\documentclass[journal,comsoc,twocolumn]{IEEEtran}

\usepackage{amsmath,amsfonts,amssymb,latexsym,verbatim, graphicx,color}
\usepackage[noadjust]{cite}

\newcommand{\constant}{near-constant}  % no caps
\newcommand{\CONSTANT}{Near-Constant}  % two caps

\newcommand{\proofsubbit}[1]{\subsubsection{#1}} % was \textbf{#1}

  \newtheorem{theorem}{Theorem}

  \newtheorem{lemma}{Lemma}

  \newtheorem{definition}{Definition}

\newcommand{\proofbit}[1]{\subsection{#1}}
\newcommand{\ee}{\mathrm{e}}
\newcommand{\bigmid}{\;\big|\;}
\newcommand{\ep}{{\mathbb {E}}}
\newcommand{\II}{{\mathbb {I}}}
\newcommand{\pr}{{\mathbb {P}}}

\newcommand{\Z}{{\mathbb{Z}}}

\newcommand{\blah}[1]{}

\newcommand{\diy}{\begin{displaystyle}}
\newcommand{\eiy}{\end{displaystyle}}
\newcommand{\suc}[1]{\pr^{{\rm #1}}({\rm suc})}
\newcommand{\err}[1]{\pr^{{\rm #1}}({\rm err})}
\newcommand{\stir}[2]{\begin{Bmatrix} #1 \\ #2 \end{Bmatrix}}
\newcommand{\smallstir}[2]{\left\{\! \begin{smallmatrix} #1 \\ #2 \end{smallmatrix}\! \right\}}
\newcommand{\bin}{{\rm Bin}}
\newcommand{\K}{\mathcal{K}}
\newcommand{\JJ}{\mathcal{J}}
\newcommand{\LL}{\mathcal{L}}
\newcommand{\PD}{\mathcal{PD}}
\newcommand{\MM}{\mathcal{M}}
\newcommand{\Ki}{\mathcal{K}^{(i)}}
\newcommand{\WW}[1]{W^{\left( #1 \right)}}
\newcommand{\WKi}{\WW{\K \setminus i}}
\newcommand{\WKij}{\WW{\K \setminus i,j}}
\newcommand{\WK}{\WW{\K}}
\newcommand{\mat}[1]{\mathsf{#1}}
\newcommand{\vect}[1]{\mathbf{#1}}
\newcommand{\wh}[1]{\widehat{#1}}
\newcommand{\gammaDD}{\gamma_{\rm DD}}
\newcommand{\gammaCOMP}{\gamma_{\rm COMP}}

\newcommand{\eb}{\mathrm{e}^{\,\beta}}
\newcommand{\com}{\mathsf{c}}

\newcommand{\rev}[1]{#1} %%% Turns new revisions blue

\title{Performance of Group Testing Algorithms \\ With \CONSTANT\ Tests-per-Item}
\author{
\IEEEauthorblockN{Oliver Johnson, Matthew Aldridge, and Jonathan Scarlett}
\thanks{O.~Johnson is with the School of Mathematics, 
University of Bristol,  University Walk, 
Bristol, BS8 1TW, UK. 
Email: maotj@bris.ac.uk. M.~Aldridge is with the Department of Mathematical Sciences, 
University of Bath, 
Claverton Down,  
Bath, BA2 7AY, UK, and the Heilbronn Institute for Mathematical Research, Bristol, UK.
Email: m.aldridge@bath.ac.uk. J.~Scarlett is with the
Departments of Computer Science and Mathematics,  National University of Singapore, 117417. 
Email: scarlett@comp.nus.edu.sg.

This work was presented in part at the 2016 IEEE International Symposium on Information Theory \cite{johnsonc14}.}}
\date{\today}

\begin{document}

\maketitle 

\begin{abstract} 
We consider the nonadaptive group testing  with $N$ items, of which $K = \Theta(N^\theta)$ are defective.
\rev{We study a test design in which each item appears in nearly the same number of tests. For each item, we independently pick $L$ tests uniformly at random with replacement, and place the item in those tests.
%We study test designs in which each item appears in a \rev{fixed} number of tests chosen uniformly at random with replacement, so that the testing matrix has near-constant column weights.
We analyse the performance of these designs with simple and practical decoding algorithms in a range of sparsity regimes, %corresponding to $\theta \in (0,1)$,
and show that the performance is consistently improved in comparison with standard Bernoulli designs. 
We show that our new design requires $23\%$ fewer tests than a Bernoulli design when paired with the simple decoding algorithms known as COMP and DD. This gives the best known nonadaptive group testing performance for $\theta > 0.43$, and the best proven performance with a practical decoding algorithm for all $\theta \in (0,1)$.
We also give a converse result showing that the DD algorithm is optimal for these designs when $\theta > 1/2$.
}
%First, we show that the very simple COMP algorithm requires $23\%$ fewer tests with a \constant\ column weight design than with a Bernoulli design., which for $\theta > 0.77$ is fewer tests than required even for optimal algorithms with Bernoulli designs.
%Second, we show that the practical DD algorithm also requires $23\%$ fewer tests with a \constant\ column weight design than with a Bernoulli design, which for $\theta > 0.43$ is fewer tests than required even for optimal algorithms with Bernoulli designs, and for all $\theta \in (0,1)$ is fewer tests than the best proven results for practical algorithms with Bernoulli designs.
%Third, we give an upper bound on the performance of \constant\ column weight designs regardless of the decoding algorithm, showing that DD is optimal for this design when $\theta \geq 1/2$.
We complement our theoretical results with simulations that show a notable improvement over Bernoulli designs in both sparse and dense regimes.
\end{abstract}

%{\bf TITLE AND ELSEWHERE: NEAR-CONSTANT??}

\section{Introduction and defintions}

\rev{In group testing, there is a population of items, some of which are `defective' in some sense. We test subsets of items called `pools'. In the standard noiseless case we consider in this paper, a test outcome is negative if every item the the pool is nondefective, and is positive if at least one item is defective. Through many such pooled tests, we hope to be able to accurately estimate which items are defective.}

The group testing problem was introduced by Dorfman \cite{dorfman}, as described in \cite[Ch.~1.1]{du}. While a wide variety of problem setups have been considered, they all share common features, and can be considered in a wider class of sparse inference problems including compressed sensing \cite{aksoylar}.
 Group testing has been applied in a wide variety of contexts, including biology \cite{chen10,gastwirth, mourad, thompson,walter},
anomaly detection in networks \cite{goodrich, lo}, signal processing and data analysis \cite{emad, gilbert}, and communications
\cite{varanasi, wolf, wu5} -- although this list is far from exhaustive.

In this paper, we prove rigorous performance bounds for the nonadaptive noiseless group testing problem.  \rev{`Nonadaptive' means that the make-up of every test pool is decided on advance, so tests can be performed in parallel.} In the common Bernoulli design, each item is placed in each test independently with some fixed probability $p$. \rev{We instead consider an alternative test design that we call the `near-constant column weight' design. Here, independently for each item, we choose $L$ tests uniformly at random with replacement and place the item in those tests.} We pair our design with practical algorithms for detecting the defective items.  Using both rigorous asymptotic results and experimental simulations, we shall see that we can accurately detect the defective items with considerably fewer tests than the Bernoulli design.

We proceed by formalizing the problem and fixing some notation. We have a large number of items $N$, of which $K$ are defective. 
We assume that the defective items are rare, with $K = o(N)$ as $N \to \infty$; moreover, for concreteness we 
follow \cite{johnsonc14, scarlett, scarlett2} by taking $K = \Theta(N^{\theta})$ for some fixed parameter $\theta \in (0,1)$. 
We follow the `combinatorial model' and
 suppose that $\K$, the true set of defective items, is chosen uniformly at random from the $\binom{N}{K}$ sets of this size.

We perform a sequence of nonadaptive tests to form an estimate $\wh{\K}$ of $\K$, and study the tradeoff between maximising the success probability $\pr( \wh{\K} = \K)$ and minimising the number of tests $T$.
We could simply take $T = N$, and test each item one by one. However, Dorfman's key insight \cite{dorfman} is that since the problem is sparse, in the sense that $K \ll N$, each test has a negative outcome with high probability, so these tests are not optimally informative. A better procedure  considers a series of pools of items that are tested together, where the outcome of each test is positive if and only if it contains at least one defective item.

\rev{A group testing procedure requires two parts. First, a \emph{test design} describes which items will be placed in which testing pools. Second, a \emph{decoding algorithm} uses the results of these tests to estimate which items are defective.}

\begin{definition} \label{def:test}
We  represent the testing pools by a (possibly random) binary matrix $\mat{X} \in \{0,1\}^{T\times N}$, where $x_{ti} = 1$ if test $t$ includes item $i$ and $x_{ti} = 0$ otherwise.
The rows of $\mat{X}$  correspond to tests, and the columns correspond to items.
\end{definition}

\begin{definition}
\rev{We consider the standard \emph{noiseless group testing model.}} The outcomes of each test are represented by a binary vector $\vect y = (y_t) \in \{0,1\}^T$, where 
a positive outcome $y_t = 1$ occurs if $x_{ti} = 1$ for some $i \in \K$, which is if the test contains a defective item.  A negative outcome $y_t=0$ occurs otherwise.
\end{definition}

\rev{A commonly used test design is the Bernoulli design -- see, for example, \cite{aldridge4, aldridge3, johnson33, atia, chan, scarlett, scarlett2}.}

\begin{definition} \label{def:bern}
\rev{We define the \emph{Bernoulli testing design} as having a testing matrix $\mat{X}$ in which each entry $x_{ti}$ is independently set to be $1$ with probability $p$ and $0$ otherwise, for some fixed parameter $p \in (0,1)$.}
\end{definition}

\rev{In this paper, we will demonstrate that better performance can be achieved by using a design we call the \constant\ column weight design.}

\begin{definition}%[Constant column weight designs]
\label{def:cc}
\rev{We define the \emph{\constant\ column weight testing design} as having a testing matrix $\mat{X}$ in which $L$ entries of each column of  are selected uniformly at random with replacement
and set to $1$, with independence between columns. The remaining entries of $\mat{X}$ are set to $0$. We set $L = \nu T/K$ for some parameter $\nu > 0$.}
\end{definition}

\rev{
We now need an algorithm to produce an estimate of the defective set. \rev{In analogy with channel coding, we can think of the defective set $\K$ as a `message' to be decoded from the `signal' $\mathbf y$, so we refer to such an algorithm as a `decoding algorithm'. (For more on connections between group testing and channel coding, see, for example, \cite{chen10,malyutov,atia,johnsonc10,aksoylar,scarlett}.)}}

\begin{definition}
We estimate the defective set by $\wh\K = \wh\K(\mat X, \vect y)$, and define the (average) success probability
  \[  \suc{} = \frac{1}{\binom NK} \sum_{|\K| = K} \pr(\wh\K = \K), \]
where the probability is over the random test design $\mat X$.
\end{definition}

\rev{We will demonstrate the superiority of our new design with two simple decoding algorithms. We define the algorithms here, but postpone detailed discussion to Section \ref{sec:algorithms}.}
\rev{First, the COMP (Combinatorial Orthogonal Matching Pursuit) algorithm is a very simple algorithm based on the fact that every item in a negative test is definitely nondefective.}

\begin{definition} \label{def:comp} \rev{The COMP algorithm is given as follows:
\begin{enumerate}
\item Mark each item that appears in a negative test as non-defective, and refer to every other item as a Possible Defective (PD) -- we write $\PD$ for the set of such items.
\item Mark every item in $\PD$ as defective.
\end{enumerate}}
\end{definition}

\rev{Second, the DD (Definite Defectives) algorithm builds on COMP to find items we can be certain are defective.}

\begin{definition}\label{def:dd}  \rev{The DD algorithm is given as follows:
\begin{enumerate}
\item Mark each item that appears in a negative test as non-defective, and refer
to every other item as a Possible Defective (PD).
\item For each positive test that contains a single Possible Defective item, mark that item as defective.
\item Mark all remaining items as non-defective.
\end{enumerate}}
\end{definition}

\rev{The main results of this paper concern rigorous bounds on the performance of the \constant\ column weight design with various decoding algorithms. We are interested in how many tests are required for the success probability to tend to $1$ as $N$ gets large. Specifically, we show the following:
\begin{itemize}
\item The COMP algorithm requires $23\%$ fewer tests with a \constant\ column weight design than with a Bernoulli design, which for $\theta > 0.77$ is fewer tests than required even for optimal algorithms with Bernoulli designs. (Theorem \ref{thm:previous1})
\item The DD algorithm also requires $23\%$ fewer tests with a \constant\ column weight design than with a Bernoulli design, which for $\theta > 0.43$ is fewer tests than required even for optimal algorithms with Bernoulli designs, and for all $\theta \in (0,1)$ is fewer tests than the best proven results for practical algorithms with Bernoulli designs. (Theorem \ref{thm:ddrate})
\item We give an upper bound on the performance of \constant\ column weight designs regardless of the decoding algorithm, showing that DD is optimal for this design when $\theta \geq 1/2$. (Theorem \ref{thm:previous2})
\item We complement our rigorous theoretical results with simulations that show a notable improvement over Bernoulli designs in both sparse and dense regimes. (Subsection \ref{sec:sims})
\end{itemize}
}

The structure of the remainder of the paper is as follows.  \rev{In Section \ref{sec:moredefs}, we define the rate of group testing (Subsection \ref{sec:rate}), formally state our main results of the paper (Subsection \ref{sec:results}), provide simulation results to illustrate the improved performance of our test design (Subsection \ref{sec:sims}), and briefly discuss some related work (Subsection \ref{sec:related}).} In Section \ref{sec:algorithms}, we describe the main decoding algorithms used \rev{in more detail} and introduce some key quantities that control their performance. In Section \ref{sec:maintheorem}, we deduce the main theorems of the paper, with proofs of some techinical results given the appendices.

\section{Further definitions and main results} \label{sec:moredefs}

\subsection{The rate of group testing} \label{sec:rate}

In this paper, we focus on nonadaptive designs, where the entire matrix $\mat X$ is fixed in advance of the tests.
In the adaptive case (where the members of each test are chosen using the outcomes of the previous tests), Hwang's generalised binary splitting algorithm \cite{hwang} recovers the defective set $\K$ using $\log_2 \binom{N}{K} + O(K)$ tests. This can be seen to be essentially optimal by a standard argument based on Fano's inequality (see for example \cite{chan}), a strengthened version of which \cite{johnsonc10} implies that any algorithm using $T$ tests has success probability bounded above by
\begin{equation} \label{eq:probuniv}
\suc{} \leq \frac{2^T}{\binom{N}{K}}.
\end{equation}

This means that any algorithm with success probability $\suc{}$  tending to 1 requires at least 
\begin{equation} \label{eq:binasymp}
T =\log_2 \binom{N}{K} \sim K \log_2\frac{N}{K} \sim (1-\theta) K \log_2  N 
\end{equation} 
tests, \rev{where $f(N) \sim g(N)$ means that $\lim_{N \to \infty} f(N)/g(N) = 1$}. (See \cite[Lemma 25]{johnson33} for details of the asymptotic behaviour of the binomial coefficient.) This motivates the following definition \cite{johnson33} of the rate of an algorithm.

\begin{definition} \label{def:rate}
For any algorithm using $T$ tests,  we define the \emph{rate} to be
\begin{equation}  \label{eq:ratedef} \frac{ \log_2 \binom{N}{K}}{T}.\end{equation}
Given a random matrix design, we say that $R$ is an \emph{achievable rate} if for any $\epsilon > 0$, there exists a group testing algorithm with rate \rev{converging to} $R$ and success probability at least $1 - \epsilon$ for $N$ sufficiently large.
\rev{We adopt the terminology \emph{maximum achievable rate} when referring to a given design (e.g., Bernoulli) and/or decoding rule (e.g., COMP).}
\end{definition}

Intuitively, one can think of the rate as being the number of bits of information learned per test when the recovery is successful.  

In this language, the result of \cite{hwang} shows that, for all $\theta \in (0,1)$, adaptive group testing has an achievable rate of $R=1$ in the regime $K = \Theta(N^{\theta})$ and is therefore optimal, since by \eqref{eq:probuniv}, no algorithm can learn more than $1$ bit per test. It is an interesting question to consider whether there exists a matrix design and an algorithm with achievable rate $R=1$ in the nonadaptive case.
It appears to be difficult even to design a class of matrices  with non-zero achievable rate using combinatorial constructions (see \cite{du,malyutov} for reviews of the extensive literature on this subject, with key early contributions coming from \cite{dyachkov2, dyachkov3, malyutov5}). Hence, much recent work on nonadaptive group testing has considered Bernoulli designs (see Definition \ref{def:bern}).  The maximum achievable rate is known exactly for such designs, as stated in the following.

\begin{theorem}\label{thm:berncap}
The  maximum achievable rate for Bernoulli nonadaptive group testing with $K = \Theta( N^{\theta} )$ defectives, \rev{for $\theta \in [0,1)$,}  is
\begin{equation}
  	C(\theta) = \max_{\nu > 0} \min \left\{ \frac{\nu\mathrm{e}^{-\nu}}{\ln  2} \frac{1-\theta}{\theta}, \,h(\mathrm e^{-\nu}) \right\} , \label{eq:bern_rate}
\end{equation}
where $h(t) = - t \log_2 t - (1-t) \log_2 (1-t)$ is the binary entropy function.
In particular, for $\theta \leq 1/3$, the maximum achievable rate of Bernoulli designs is 1.
\end{theorem}

\rev{The direct part of Theorem \ref{thm:berncap} is due to \cite{scarlett} and the converse due to \cite{aldridge4}. (The special case $\theta = 0$ is older \cite{friedlina}.)}

The curve \eqref{eq:bern_rate} is illustrated in Figure \ref{fig:ratecomparison} below.
For $\theta \geq 1/2$, the paper \cite{johnson33} showed that
  \eqref{eq:bern_rate} is achieved by the DD algorithm described above.
However, for $\theta < 1/2$, the algorithms known to achieve the bound \eqref{eq:bern_rate} are based on maximising the likelihood or solving other difficult combinatorial problems, and cannot be considered as practical in a computational sense -- see
Section \ref{sec:feasible} for more details. For 
example, we describe the SSS algorithm in Definition \ref{def:sss} below, which achieves the bound of \cite{aldridge4}, but is impractical for large values of $N$ and $K$.

\subsection{Main results} \label{sec:results}

\rev{Our main results concern improving on Theorem \ref{thm:berncap} by using a \constant\ column weight design (Defintion \ref{def:cc}). Recall that this design has a testing matrix $\mat{X}$ in which $L = \nu T/K$ entries of each column of are selected uniformly at random with replacement
and set to $1$, with independence between columns, and the remaining entries of $\mat{X}$ are set to $0$.} \rev{The tester may choose $L$ to depend on the parameters of the group testing problem.}

%Our main results provide strict improvements on Theorem  \ref{thm:berncap} in a range of sparsity regimes, as well as strict improvements over existing practical algorithms in all scaling regimes.  To do this, we make use of the following class of test designs (see Section \ref{sec:related} for previous works considering similar designs).

%%%%% DEFINITION OF CCWD WAS HERE

%%% A paragraph that discusses constant v. near constant
%We use the word `constant' to mean that the same number $L$ of tests is chosen for each item. 

Since the tests are chosen \emph{with replacement}, some columns may actually have weight slightly less than $L$ due to the same test being picked more than once, hence we use the term `near-constant'. Since the weight of a column is the number of tests an item is in, we also consider these designs as `near-constant tests-per-item'. In a preliminary report \cite{johnsonc14}, we used the less precise terminology `constant column weight' for these same designs. \rev{Evidence from simulations and heuristic calculations suggest that truly-constant column weight designs have the same performance as the near-constant designs we consider here, but the rigorous analysis of such designs seems more difficult.}\footnote{\rev{As pointed out by a reviewer, the COMP rate in Theorem \ref{thm:previous1} can be shown to be achieved by a truly-constant column weight design with little extra difficulty. However, we have not been able to rigorously verify that the same is true for our main result, Theorem \ref{thm:ddrate}, or for the algorithm-independent converse, Theorem \ref{thm:previous2}.} \label{foot:exact_cc}}

\begin{figure*}
\centering
\includegraphics[width=0.85\textwidth]{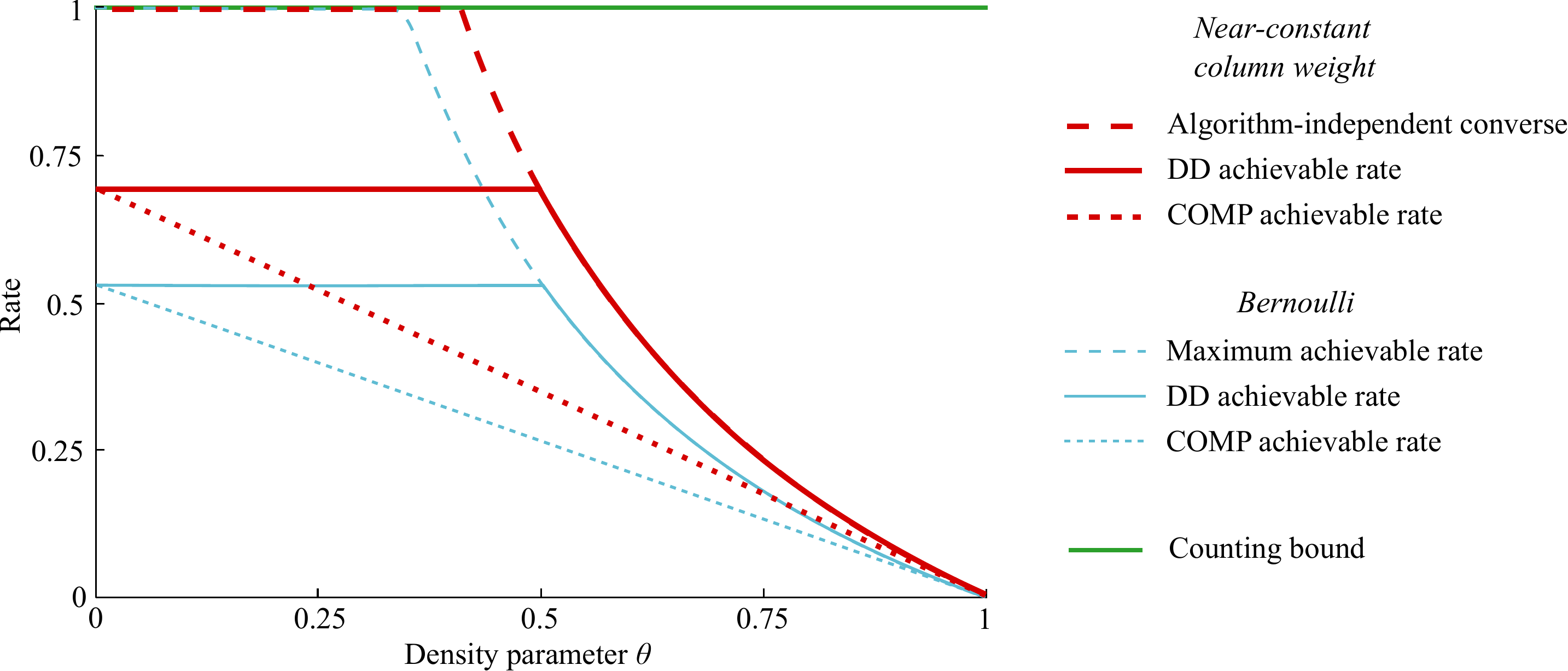}
\caption{Rates and bounds for group testing algorithms with Bernoulli designs and \constant\ column weight designs. In thick red, we plot the rate bounds for \constant\ column weight designs from Theorems \ref{thm:previous1}, \ref{thm:ddrate} and \ref{thm:previous2}. In thin blue, we plot the rate bounds for Bernoulli designs from Theorem \ref{thm:berncap}
and \cite{johnson33}. The green horizontal line represents the universal `counting bound' arising from \eqref{eq:probuniv}.
}
\label{fig:ratecomparison}
\end{figure*}

The main results of this paper are the following three theorems. The COMP and DD algorithms were defined in Defintions \ref{def:comp} and \ref{def:dd} and are discussed further in Section \ref{sec:algorithms}. The main results are proved in Section \ref{sec:maintheorem}. We illustrate these results in Figure \ref{fig:ratecomparison}, with our new rates for \constant\ column weight designs marked in thick red, and corresponding rates for Bernoulli designs marked in thin blue. 

Our first result concerns the simple and practical COMP decoding algorithm (see Defintion \ref{def:comp} and Subsection \ref{sec:COMP}), based on the fact that all items in negative tests must be negative.  %This was studied by Chan \emph{et al.}~\cite{chan} and also dates back to earlier works \cite{kautz}.

\begin{theorem} \label{thm:previous1}
Consider a \constant\ column weight design with an optimised parameter $\nu > 0$. When there are $K = \Theta( N^{\theta} )$ defectives \rev{for $\theta \in [0,1)$},
the COMP algorithm has success probability tending to $1$ if
$T \geq (1+\epsilon) T^{\mathrm{COMP}}$, and tending to $0$ if $T \leq (1-\epsilon) T^{\mathrm{COMP}}$. 
Here
\[ T^{\mathrm{COMP}} = \frac{1}{\ln  2} K \log_2  N . \]
Hence, the COMP algorithm has maximum achievable rate $\ln  2 \,(1-\theta)$ \rev{for all $\theta \in [0,1)$}.
\end{theorem}

This rate 
  \[ \ln  2 \,(1-\theta) \,\rev{\approx}\, 0.693(1-\theta) \]
is an improvement by \rev{$30.6\%$} on the rate
  \[ \frac{1}{\mathrm e \ln  2} (1-\theta) \,\rev{\approx}\, 0.531 (1-\theta) \]
for COMP with a Bernoulli design \cite{chan,aldridge4}. (An improvement in rate of \rev{$30.6\%$} corresponds to using $23.4\%$ fewer tests.) Further, for $\theta > 0.766$, Theorem \ref{thm:previous1} is an improvement on \eqref{eq:bern_rate}, meaning that in dense cases, the very simple COMP algorithm with a \constant\ column weight design beats \emph{any} decoding algorithm with a Bernoulli design (see Figure \ref{fig:ratecomparison}).

Some insight on this result can be attained by considering the conditions under which COMP succeeds.  Under the choice $\nu = \ln 2$, Bernoulli testing with probability $\nu/\rev{K}$ and a \constant\ column weight design with $L = \nu T/\rev{K}$ both result in roughly half of the tests being positive (e.g., see Lemma \ref{lem:mcdiarmid} below).  However, a given non-defective item is placed in roughly $\mathrm{Binomial}\big(\frac{T}{2}, \frac{\ln 2}{\rev{K}} \big)$ negative tests under Bernoulli testing, and $\mathrm{Binomial}\big(\frac{T \ln 2}{\rev{K}}, \frac{1}{2} \big)$ negative tests under the near-constant column weight design.  While these two distributions have the same expectation, the latter has a much smaller probability of being zero, which is the event under which COMP fails.  

It is also interesting to note that while $\nu = \ln 2$ (which is 'maximally informative' in the sense of maximising the entropy of the test outcome) optimises the rate of COMP (as well as DD below) for the near-constant column weight design, COMP \cite{chan} and DD \cite{johnson33} with Bernoulli designs are optimised with a fraction $1 - \ee^{-1} \approx 0.632$ of positive tests.

% The choice $\nu = \ln 2$ leads  to roughly a fraction $1 - \ee^{-\ln 2} = 0.5$ (see Lemma \ref{lem:mcdiarmid} below) of tests being positive, which can be thought of as `maximally informative' in the sense of maximising entropy. By comparison, COMP \cite{chan} and DD \cite{johnson33} with Bernoulli designs are optimised with a fraction $1 - \ee^{-1} = 0.632$ of positive tests.

Our second and most important result concerns the practical DD decoding algorithm (see Definition \ref{def:dd} and Subsection \ref{sec:DD}).

\begin{theorem} \label{thm:ddrate}
Consider a \constant\ column weight design with an optimized parameter $\nu > 0$. When there are $K = \Theta( N^{\theta} )$ defectives \rev{for $\theta \in (0,1)$},
the DD algorithm has success probability tending to $1$ if
\[ T \geq (1+\epsilon)\frac{1}{\ln  2} \max\left\{ K \log_2  \frac{N}{K}, K \log_2 K \right\} , \]
and hence has an achievable rate 
   \[ R = \ln  2 \, \min \left\{ 1,  \frac{1-\theta}{\theta} \right\} = \begin{cases} \ln 2 & \theta \leq \tfrac12 \\ \ln 2 \, \displaystyle\frac{1-\theta}{\theta} & \theta > \tfrac12. \end{cases} \]
\end{theorem}

This rate 
  \[ \ln  2 \, \min \left\{ 1,  \frac{1-\theta}{\theta} \right\} \rev{\approx}\, 0.693\,\min \left\{ 1,  \frac{1-\theta}{\theta} \right\} \]
is an improvement again by \rev{$30.6\%$} on the rate of
  \[ \frac{1}{\mathrm e \ln  2} \,\min \left\{ 1,  \frac{1-\theta}{\theta} \right\} \rev{\approx}\, 0.531 \,\min \left\{ 1,  \frac{1-\theta}{\theta} \right\} \]
proved by \cite{johnson33} for DD with Bernoulli designs. In fact, to our knowledge, DD with the \constant\ column weight design gives the highest proven {\em practically achievable} rate for all $\theta \in (0,1)$. Further, for $\theta > 1/(1 + \mathrm e (\ln  2)^2) \approx 0.434$, Theorem~\ref{thm:ddrate} is an improvement on \eqref{eq:bern_rate}, meaning that in this regime, the practical DD algorithm with a \constant\ column weight design beats any decoding algorithm (even impractical ones) with a Bernoulli design.

Our third result is an algorithm-independent converse, showing the maximum possible rate of any decoding algorithm with a \constant\ weight design.

\begin{theorem} \label{thm:previous2}
Consider a \constant\ column weight design, with $K = \Theta( N^{\theta} )$ defectives \rev{for $\theta \in (0,1)$}.
Regardless of the choice of $\nu > 0$, no algorithm can achieve a rate greater than
\begin{equation} \label{eq:conv}
  \min \left\{ 1 , \ln  2 \, \frac{1-\theta}{\theta}  \right\} = \begin{cases} 1 & \theta \leq \theta^* \\ \ln 2 \, \displaystyle\frac{1-\theta}{\theta} & \theta > \theta^*,\end{cases}
\end{equation}
where
  \[ \theta^* = \frac{\ln 2}{1+\ln 2} \approx 0.409 . \]
\end{theorem}

Comparing Theorems \ref{thm:ddrate} and \ref{thm:previous2}, we see that if we use a \constant\ column weight design, the DD algorithm gives the optimal performance for $\theta \geq 1/2$.

% The choice $\nu = \ln 2$ leads  to roughly a fraction $1 - \ee^{-\ln 2} = 0.5$ (see Lemma \ref{lem:mcdiarmid} below) of tests being positive, which can be thought of as `maximally informative' in the sense of maximising entropy. By comparison, COMP \cite{chan} and DD \cite{johnson33} with Bernoulli designs are optimised with a fraction $1 - \ee^{-1} = 0.632$ of positive tests.

%***** In \cite{johnsonc14}, it was shown that $\nu = \ln 2$ is optimal with respect to all of the bounds derived therein, and we will also use this value throughout the present paper. This choice ensures that each test is equally likely to be positive or negative, and can thus be thought of as being maximally informative. For brevity, we prove Theorem \ref{thm:previous2} with $\nu = \ln 2$ rather than the general case. *****

% ***** At an intuitive level, we believe the improvement over Bernoulli designs arises due to the fact that the latter can result in some items appearing in considerably fewer tests than average, meaning that it is harder for any algorithm to infer their defectivity status. *****

\begin{figure*}
\centering
\includegraphics[width=0.72\textwidth]{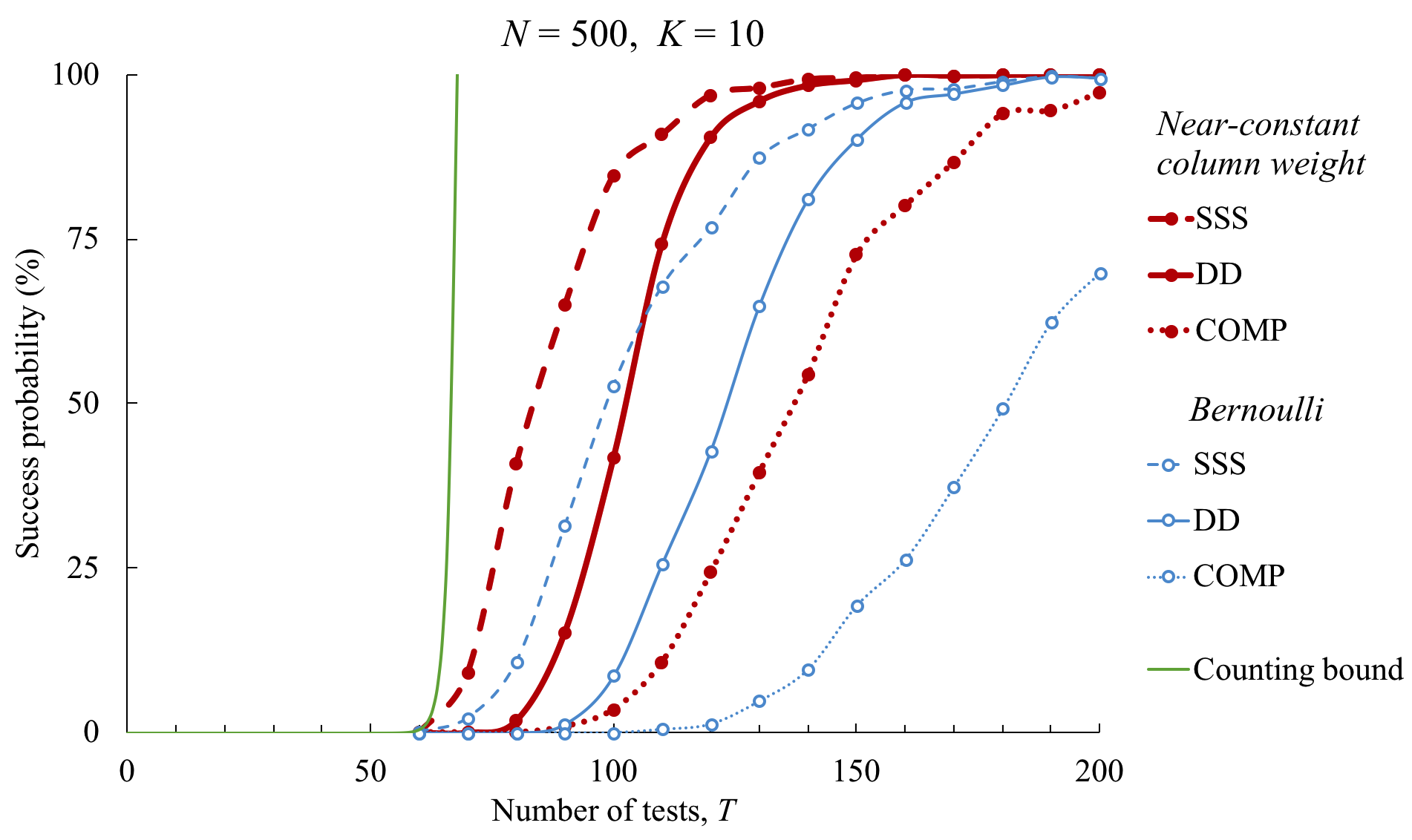}

\bigskip
\includegraphics[width=0.72\textwidth]{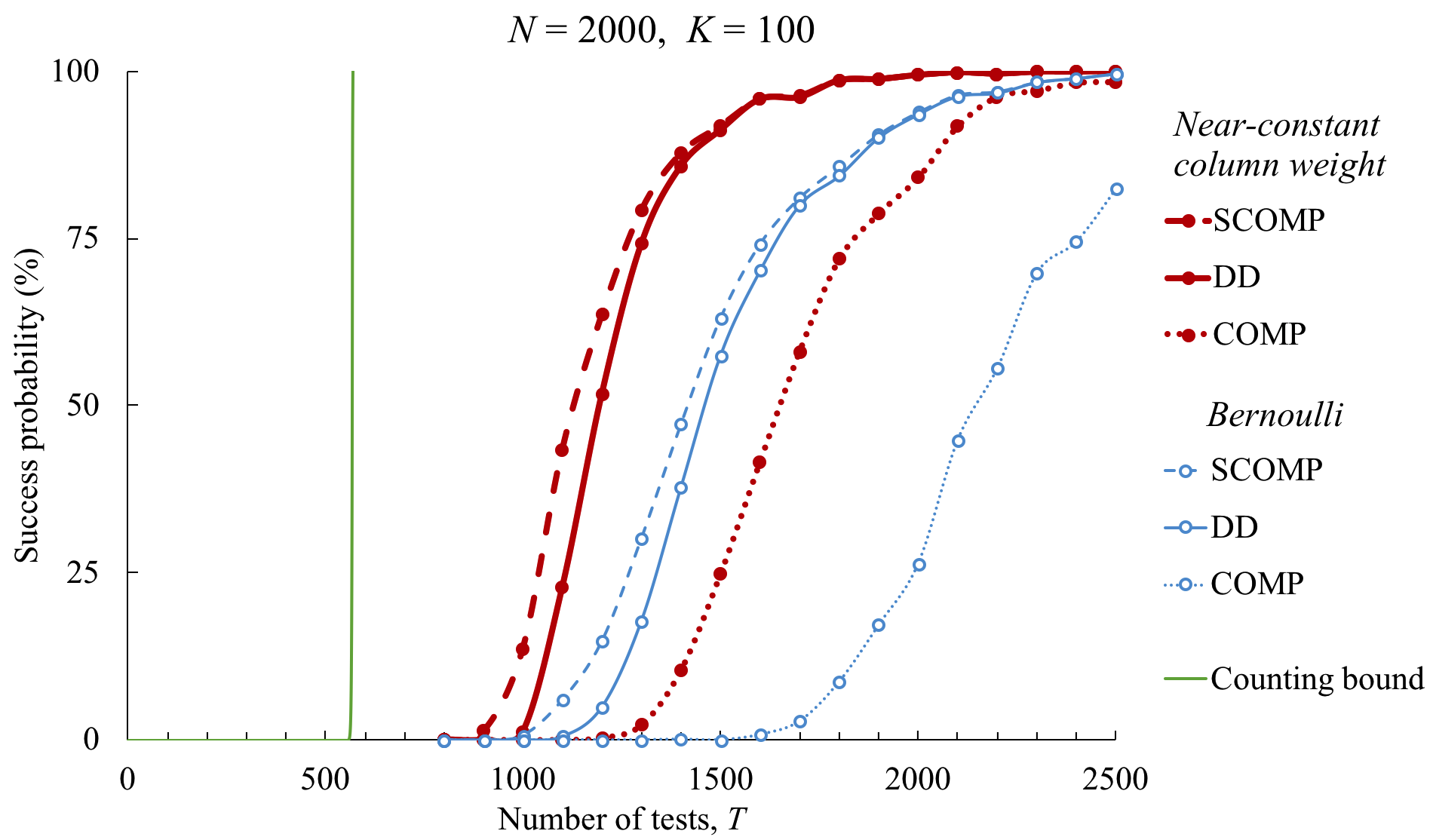}
\caption{Empirical performance (each point based on 1000 simulations) of various algorithms for both \constant\ column weight and Bernoulli
designs, in the cases $N=500$, $K = 10$ and $N=2000$, $K=100$.}
\label{fig:finite}
\end{figure*}

\subsection{Simulations} \label{sec:sims}

We complement our rigorous results on the rate, which are asymptotic as $N \to \infty$, 
with simulations that show \constant\ weight designs also improve on Bernoulli designs for finite problem sizes. In Figure \ref{fig:finite}, we illustrate the performance of these algorithms via simulations in an illustrative sparse case ($N = 500$, $K=10$) and a denser case ($N=2000$, $K=100$). For the sparse case, in addition to plotting performance of COMP and DD, we plot the performance of the SSS algorithm (see Definition \ref{def:sss}), which 
achieves the bounds of Theorem \ref{thm:berncap} \cite[Corollary 4]{aldridge3}, though is not practical for larger problems.  Because of this issue of practicality, we do not consider SSS for the denser case; instead, we plot the performance of a  related algorithm called SCOMP, which is described in \cite{johnson33}, so we omit a description in this paper for the sake of brevity.  (Essentially, it amounts to performing DD followed by greedy refinements.) Our \constant\ weight designs provide a consistent notable improvement on Bernoulli designs, particularly in the denser example.

\subsection{Related work} \label{sec:related}

While we are not aware of previous works using our exact \constant\ column weight design, closely-related designs have been proposed.  Our  key  contribution is a \emph{rigorous} analysis of such designs in the regime $K = \Theta( N^{\theta} )$, requiring several novel techniques.  In particular, we prove the achievability of rates strictly above those achieved by Bernoulli designs. 

Kautz and Singleton \cite{kautz} observed that a construction based on a concatenation of constant-weight codes gives matrices with the so-called $K$-disjunctness property (the union of any $K$ columns does not contain any other column). Such matrices give group testing designs guaranteeing that $K$ defectives can be recovered  with zero probability of error in noiseless group testing (see for example \cite[Chapter 7]{du}). However, the group testing designs resulting from the construction of \cite{kautz} require $T = O(K^2 (\log N)^2)$ tests. This is an example of the fact that the zero-error criterion requires considerably more tests than the $T = O(K \log N)$ required for the `error probability tending to zero' criterion (see Definition \ref{def:rate}) that we study here.

Similarly, other subsequent papers have proposed forms of constant or near-constant column weight designs \cite{balding1996comparative,bruno1995efficient,hwang2000random,macula1998probabilistic}, but to our knowledge, none of these works provide non-trivial achievable rates for the vanishing error probability criterion, which is the focus of the present paper.  Chan \emph{et al.} \cite{chan} considered constant {\em row} weight designs, and found no improvement over Bernoulli designs.

M\'ezard \emph{et al.} \cite{mezard} considered randomised designs with both constant row and column weights, and with constant column weights only.  The paper used heuristics from statistical physics to suggest that such designs may beat Bernoulli designs. In our notation, they suggest the maximum achievable rate of these constant weight designs may be equal to our converse bound \eqref{eq:conv} for all $\theta$. \rev{(Our Theorem \ref{thm:ddrate} rigorously proves this for $\theta \geq 1/2$ under our slightly different design.) The work of \cite{mezard} contains some non-rigorous steps; in particular, they make use of a `no short loops' assumption that is only verified for $\theta > \frac{5}{6}$ and conjectured for $\theta > \frac{2}{3}$, while experimentally being shown to {\em fail} for smaller values such as $\theta = \frac{1}{3}$.}

  % Wadayama \cite{wadayama} analysed constant row and column weight designs in the linear regime $K = cN$, and demonstrated close-to-optimal asymptotic performance for certain ratios of parameter sizes. In contrast, we consider the fundamentally different sparse regime $K = o(N)$ here.

D'yachkov {\em et al.} \cite{dyachkov4} studied list decoding with (exactly) constant column weight designs, and setting their list size to $1$ corresponds to insisting that COMP succeeds. 
However, they only considered the case that $K = O(1)$. In the limit as $K$ gets large, the rate $\ln 2$ obtained \cite[Claim 2]{dyachkov4} matches the rate for COMP given here in Theorem \ref{thm:previous1} in the limit $\theta \rightarrow 0$.

A distinct line of works has sought designs that not only require a low number of tests, but also near-optimal {\em decoding complexity} (e.g., $K \mathrm{poly}(\log N)$) \cite{Cai13,Lee15a,Ngo11,Ind10}.  However, our focus in this paper is on the required number of tests, for which the existing guarantees of such algorithms contain loose constants or extra logarithmic factors.

\section{Decoding algorithms: further details} \label{sec:algorithms}

In this section we discuss the COMP and DD algorithms in more detail, and introduce the SSS (Smallest Satisfying Set) algorithm. We discuss the conditions under which these algorithms succeed. These algorithms were previously studied in \cite{chan} and \cite{johnson33} with Bernoulli designs.   Before continuing, we present another key definition:

\begin{definition} \label{def:masked} Consider an item $i$ and a set of items $\LL$ not including $i$.
We say that item $i$ is \emph{masked} by $\LL$ if every test that includes $i$ also includes at least one member of $\LL$.
\end{definition}

\subsection{COMP algorithm} \label{sec:COMP}

Recall the COMP algorithm from Definition \ref{def:comp}:
\begin{enumerate}
\item Mark each item that appears in a negative test as non-defective, and refer to every other item as a Possible Defective (PD) -- we write $\PD$ for the set of such items.
\item Mark every item in $\PD$ as defective.
\end{enumerate}
This is based on a simple inference: Any negative test only contains non-defective items, so any item 
in a negative test can be marked as non-defective. Given
enough negative tests, we might hope to correctly infer every member of $\K^\com$ in this way.  \rev{The name COMP (Combinatorial Orthogonal Matching Pursuit) was coined in \cite{chan}, although the method itself appeared much earlier -- see, for example, \cite{kautz,malyutov,chen10,luo}.}

Clearly, the first step will not make any mistakes (every item marked as non-defective will indeed be non-defective), so errors
will only occur in the second step. As a result COMP will always estimate $\K$ by a set $\wh{\K}_{\rm COMP}$ with
$\K \subseteq \wh{\K}_{\rm COMP}$.

As in \cite{johnson33}, a quantity of particular interest is $G := |\PD \setminus \K| = |\PD| - K$, the number of non-defective items masked by the defective set $\K$. So $G$ is the number of non-defective items that do not appear in any negative test.
It is clear that COMP succeeds (recovers the defective set exactly) if and only if $G=0$, so
that 
\begin{equation} \label{eq:compsuc} \suc{COMP} = \pr(G=0).\end{equation} 
We use this in the proof of Theorem \ref{thm:previous1} in Section \ref{sec:maintheorem}.

\subsection{DD algorithm} \label{sec:DD}

Recall the DD algorithm from Definition \ref{def:dd}, which builds on COMP to find items that are definitely defective:
\begin{enumerate}
\item Mark each item that appears in a negative test as non-defective, and refer
to every other item as a Possible Defective (PD).
\item For each positive test that contains a single Possible Defective item, mark that item as defective.
\item Mark all remaining items as non-defective.
\end{enumerate}
The performance of the DD algorithm with Bernoulli designs was studied in detail by Aldridge, Baldassini and Johnson \cite{johnson33}.

Again, the first step will not make any mistakes, and since every positive test must contain at least one defective item, the second 
step is also certainly correct. Hence, any errors due to DD come from marking a true defective as non-defective in the third step, 
meaning that the estimate $\wh{\K}_{\rm DD}$ satisfies $\wh{\K}_{\rm DD} \subseteq \K$. The choice to mark all 
remaining items as non-defective is motivated by the sparsity of the problem, since {\em a priori} an item is much less likely to be defective than non-defective.

We analyse DD rigorously in Section \ref{sec:maintheorem}, using the following notation, used in \cite{johnson33}.
% and illustrated in Figure \ref{fig:itemrvs}.
 For each $i \in \K$,  we write:
\begin{itemize}
	\item $M_i$ for the number of tests containing defective item  $i$ and no other defective;
    \item $L_i$ for the number of tests containing defective item $i$ and no other possible defective item (no other member of $\PD$).
\end{itemize}
In the terminology of Definition \ref{def:masked}, we see that DD succeeds if and
only if no defective item $i \in \K$ is masked by $\PD \setminus \{i \}$. Further, since item $i$ is masked by $\PD \setminus \{i \}$ if and only if 
$L_i= 0$, we can write
\begin{equation} \label{eq:ddsuc} 
\suc{DD} = 1 - \pr \left( \bigcup_{i \in \K} \{ L_i 
\rev{=}  0 \} \right).\end{equation}

For a given defective item $i \in \K$, we write $\Ki = \K \setminus \{ i \}$ for the set of defectives with $i$ removed.
For a given set $\MM$, we write $W^{(\MM)}$ for the total  number of tests containing at least one item from $\MM$. The
random variable $W^{\K\setminus\{i\}}$ (the total  number of tests containing at least one item in $\Ki$), henceforth denoted by $\WKi$, will be of particular interest.

To understand the distributions of these quantities,
%illustrated in Figure \ref{fig:itemrvs}, 
 it is helpful to think of the process by which elements of the columns are sampled as a \emph{coupon collector} problem, where each coupon corresponds to one of the $T$ tests.  For a single item, $\WW{\{ i \}}$ is the number of distinct coupons selected when $L$ coupons are chosen uniformly at random from a population of $T$ coupons. In general, for a set $\MM$ of size $M$, 
the independence of distinct columns means that $\WW{\MM}$ is the number of distinct coupons collected when choosing $ML$ coupons uniformly at random from a population of $T$ coupons.

Hence, as described in more detail in Section \ref{sec:maintheorem}, we can first give a concentration of measure result for $\WKi$ (see Lemma \ref{lem:mcdiarmid}), then characterise the distribution of $M_i$ given $\WKi$ (see Lemma \ref{prop:wdef}). Following this, we can state the distribution of $G$ conditioned on $\WK = \WKi + M_i$ (see Lemma \ref{lem:gdef}), and finally deduce the distribution of $L_i$ conditioned on $G$ and $\WK$ (see Lemma \ref{lem:plkzero}). This allows us to deduce bounds on \eqref{eq:ddsuc}.

\subsection{SSS algorithm}

We describe one more algorithm, which we call the SSS (Smallest Satisfying Set) algorithm, following \cite{johnson33}. This algorithm is not directly mentioned in the statement of our main results, but its analysis will be important in proving Theorem \ref{thm:previous2}.

\begin{definition} \label{def:sss}
We say that a putative defective set $\JJ$ is \emph{satisfying} if: 
\begin{enumerate}
\item No negative test contains a member of $\JJ$.
\item Every positive test contains at least one member of $\JJ$.
\end{enumerate}
The SSS algorithm simply finds the smallest satisfying set (breaking ties arbitrarily), and takes that as the estimate $\wh{\K}_{\rm SSS}$.
\end{definition}

Note that the true defective set $\K$ is certainly a satisfying set, and hence SSS is guaranteed to return a set of no larger size, so $| \wh{\K}_{\rm SSS}| \leq | \K|$. However, it may not be the case that $ \wh{\K}_{\rm SSS} \subseteq \K$. We can identify a particular failure event for SSS: If a defective item $i \in \K$ is masked by the other defective items $\K \setminus \{i \}$ (in the sense of Definition \ref{def:masked}) then $\K \setminus \{i \}$ will be a smaller satisfying set, so SSS is certain to fail. 

Hence, writing $A_i$ for the event that item $i$ is masked by $\K \setminus \{ i\}$, we can use the Bonferroni inequality to obtain a lower bound on the SSS error probability $\err{SSS}$ of the form 
\begin{equation}  \label{eq:SSSfailure} 
\err{SSS} \geq
\pr \left( \bigcup_{i \in \K}  A_i
 \right) \geq \sum_{i \in \K} \pr(A_i) - \frac{1}{2} \sum_{i \neq j \in \K} \pr
 \left( A_i \cap
 A_j \right).
\end{equation}
This serves as a starting point for upper bounding the rate of the SSS algorithm, which in turn will be used to infer our general converse (Theorem \ref{thm:previous2}).

\subsection{Note on practical feasibility} \label{sec:feasible}

We refer to COMP and DD as `practical' algorithms, since they can be implemented with low run-time and storage. For example, COMP simply requires us to take one pass through the test matrix and outcomes, requiring  no more than $O(N)$ storage beyond the matrix itself, and $O(T N)$ runtime. Similarly, DD builds on COMP, requiring two passes through the test matrix and outcomes and can be performed with the same amount of storage and runtime. 

In contrast, we can interpret SSS as an integer programming problem, meaning that it is unlikely to be practical to run for large problems.  We think of it as the `best possible' algorithm without knowing $K$, and use a rigorous form of this statement \cite{aldridge4} to obtain algorithm-independent performance bounds. Note that although the SSS algorithm may be considered to be infeasible in practice, the papers \cite{malioutov2,aldridge-dd} show that a relaxation of the integer programming problem to the real numbers can give good performance. 

Furthermore, the decoding algorithms we consider here do not require exact, or even approximate, knowledge of $K$. This is in contrast to the optimal maximum likelihood decoder of \cite{scarlett}, which requires the exact value of $K$. Note, however, that the optimal choice of the parameter $\nu = (\ln 2)T/K$ in the \emph{design} stage does require knowing $K$.

\section{Proofs of main results} \label{sec:maintheorem}

The main goal of this section is to prove our achievable rate for the DD algorithm (Theorem \ref{thm:ddrate}). Along the way, we will also prove the COMP rate (Theorem \ref{thm:previous1}) and the algorithm-independent upper bound (Theorem \ref{thm:previous2}); the former will essentially come `for free', though the latter will require non-trivial additional effort.

\proofbit{Concentration of $\WW{\MM}$}

Recall that $\WW{\MM}$ corresponds to the total number of tests in which items from $\MM$ are placed. The following lemma shows that this quantity concentrates around its mean.

\begin{lemma} \label{lem:mcdiarmid}
Let $M = |\mathcal M|$, and fix the constants $\alpha > 0$ and $\epsilon \in (0,1)$. When making
$L M = \alpha T$ draws with replacement from a total of $T$ coupons,
the total number of distinct coupons $\WW{\MM}$
satisfies
\begin{equation} \label{eq:mcdiarmid}
 \pr \big( \big| \WW{\MM} - (1-\ee^{-\alpha}) T \big| \geq \rev{\delta} \big) \leq 2 \exp \left( -\frac{\rev{\delta^2}}{\alpha \rev{T}} \right)
\end{equation}
for $T$ sufficiently large.
\end{lemma}

\begin{IEEEproof}
We first characterise the expectation of $\WW{\MM}$, and then show concentration about that expectation.  By the linearity of expectation, we have
  \begin{align*}
    \mathbb E \WW{\MM}
      &= \sum_{j = 1}^T \pr(\text{coupon $j$ in first $LM$ selections}) \\
      &= \sum_{j = 1}^T \left(1 - \left(1 - \frac1T\right)^{LM} \right) \\
      &= \left(1 - \left(1 - \frac1T\right)^{\alpha T} \right) T.
  \end{align*}
It follows that $\mathbb E \WW{\MM} = (1-\ee^{-\alpha}) T + o(T)$ as $T \to \infty$.

To establish concentration about the mean, we use McDiarmid's inequality \cite{mcdiarmid}, which characterises the concentration of functions of independent random variables when the bounded difference property is satisfied.  Write $Y_1, Y_2, \dots, Y_c$ for the labels of the selected coupons, and
  $W(c) = f(Y_1, Y_2, \dots, Y_c)$
for the number of distinct coupons. Note that here we have the bounded difference property, in that
  \[ \big| f(Y_1, \dots, Y_j, \dots, Y_c)
         - f(Y_1, \dots, \hat Y_j, \dots, Y_c) \big| \leq 1 \]
for any $j$, $Y_1, \dots, Y_c,$ and $\hat Y_j$, since the largest difference we can make is swapping a distinct coupon $Y_j$ for a non-distinct one $\hat Y_j$, or vice versa.
McDiarmid's inequality \cite{mcdiarmid} gives that
  \[
    \pr \big( \big| f(Y_1, \dots, Y_c)
      - \mathbb E f(Y_1, \dots, Y_c) \big| \geq \delta \big)
      \leq 2 \exp \left( -\frac{2 \delta^2}{c} \right) .
  \]
Setting  $c = \alpha T$ gives the desired result; we crudely remove the factor of $2$ from the exponent to account for the fact that we are considering deviations from the asymptotic value of the mean of $\WW{\MM}$ rather than the exact value, which amounts to the replacement of $\alpha$ by $\alpha(1+o(1))$.
\end{IEEEproof} 

\proofbit{Proof of the algorithm-independent converse}

The above concentration result plays an important role in the proof of Theorem \ref{thm:previous2}.

\begin{IEEEproof}[Proof of Theorem \ref{thm:previous2}]
\rev{We divide the proof into three steps. First, we begin with an overview of some preliminary results that will be used throughout the proof; second, we bound the error probability of the SSS algorithm; and third, we bound a key quantity that arises in the proof.}
    
\proofsubbit{Preliminaries}
\rev{The initial steps follow the proof of a similar result for Bernoulli testing in \cite{aldridge4}.
As shown there, if $\err{SSS} + \err{COMP} > 1 + \epsilon$ for some $\epsilon > 0$ that remains bounded away from zero as $N \to \infty$, then the error probability is also bounded away from zero for an arbitrary algorithm.  We know the condition under which $\err{COMP} \to 1$ from Theorem \ref{thm:previous1} (which will be proved later), and it is easy to see that the corresponding bound is weaker than that of Theorem \ref{thm:previous2}, since}
    \[ 1-\theta \leq  \min \left\{1,\frac{1-\theta}{\theta} \right\}. \]
\rev{Hence, it suffices to show that the error probability of the SSS algorithm is bounded away from zero; we do so in the remainder of the proof.}

The upper bound of $1$ on the rate is well-known for arbitrary test designs (this follows from \eqref{eq:probuniv}, for example), so we only need to obtain the other term in \eqref{eq:conv}.  To do so, we claim that it suffices to show that for any choice of $\nu > 0$ (such that $L = \nu T/K$) the error probability is bounded away from zero for some $T$ satisfying
\begin{equation}
    T = \frac{ K \ln K}{ -\nu \ln(1-\ee^{-\nu}) } (1+o(1)). \label{eq:T_choice}
\end{equation}
To see that it suffices to choose $T$ in this way, first note that $-\nu \ln(1-\ee^{-\nu})$ attains its maximum of $(\ln 2)^2$ at $\nu = \ln 2$, in which case the rate corresponding to \eqref{eq:T_choice} is $\ln2 \frac{1-\theta}{\theta}$, as required.  For other choices of $\nu$, the choice \eqref{eq:T_choice} corresponds to more tests than dictated by the rate of Theorem \ref{thm:previous2}, but this is allowed for the purpose of proving a converse, since additional tests can never hurt the SSS algorithm.

We can also assume that $\nu$ is constant, since it is straightforward to verify that the cases $\nu \to 0$ or $\nu \to \infty$ fail to even yield the correct scaling $T = \Theta(K \ln N)$.  This is because, in such cases, the probability of a given test being positive tends to either $0$ or $1$, and hence the entropy of the test vanishes.

%***** We thus choose $T$ according to the other term, setting $T = \gammaSSS \theta K \ln N$ with $\gammaSSS = -(1-\epsilon)/\nu \ln (1 - \ee^{-\nu)}$.  Hence, $L = \nu T/K = (1-\epsilon) \theta \ln N/(1 - \ee^{-\nu)}$ (THIS IS A GUESS). We want to show that the error probability is bounded away from $0$.*****

Finally, we note the following concentration result: Lemma \ref{lem:mcdiarmid} above shows that for both $M = K-2$ and $M=K-1$, choosing $\alpha = L M/T = \nu M/K \to \nu$ reveals that $\WW{\MM}$ is exponentially concentrated around $(1 - \ee^{-\nu})T$.  In particular, there exists a constant $c_0 > 0$ such that
\begin{equation}
    \Pr\Big( \big| \WW{\MM} - (1 - \ee^{-\nu})T \big| \rev{ \geq } \sqrt{c_0 T \ln T} \Big) \le \frac{1}{T^3} \label{eq:sss_conc}
\end{equation}
for sufficiently large $T$.

\proofsubbit{Bounding the error probability of SSS}
\rev{We start with the lower bound on the error probability given in \eqref{eq:SSSfailure}, which we repeat here for convenience:
\begin{equation}  \label{eq:SSSfailure_rep} 
    \err{SSS} \geq
    \pr \left( \bigcup_{i \in \K}  A_i
    \right) \geq \sum_{i \in \K} \pr(A_i) - \frac{1}{2} \sum_{i \neq j \in \K} \pr
    \left( A_i \cap
    A_j \right).
\end{equation}
We will show that for any constant $\nu > 0$, the right-hand side of \eqref{eq:SSSfailure_rep} is bounded away from zero as $N \to \infty$ under some number of tests $T$ satisfying \eqref{eq:T_choice}.  We begin by bounding the two terms on the right-hand side of \eqref{eq:SSSfailure_rep}.}

\begin{lemma} \label{lem:sss_probs}
    \rev{
    Under the preceding definitions, and under a near-constant column weight design with parameter $\nu > 0$, we have for any constants $c_1,c_2 > 0$ and $\epsilon_1 \in (0,1)$ that
    \begin{align}
        \qquad K \pr(A_i) &\ge K c_1^L \,\pr\big( \WKi \geq T c_1\big) \label{eq:term1} \\
        \binom{K}{2} \pr(A_i \cap A_j)
        &\le \frac{K^2}{2(1-\epsilon_1)}\, \bigg(c_2 + \frac{\nu}{K}\bigg)^{2L} \pr\big(  Tc_1 \leq \WKij \leq T c_2\big) \notag \\
        & \quad + \binom{K}{2} \pr\big( \WKij < T c_1\big)  + \binom{K}{2} \pr\big( \WKij > T c_2\big) \label{eq:term2}
    \end{align}
    when $N$ is sufficiently large.
    }
\end{lemma} \medskip
\rev{The idea of the proof is to lower bound $\pr(A_i)$ by restricting attention to the event $\WKi \geq T c_1$ and applying counting arguments, and to upper bound $\pr(A_i \cap A_j)$ by one when the suitable bounds on $\WKij$ fail to hold, while upper bounding it using counting arguments otherwise.  The details are given in Appendix \ref{sec:pf_sss_probs}.}

From \eqref{eq:sss_conc}, if we choose \rev{
\begin{align}
  c_1 &= 1 - \mathrm{e}^{-\nu} - \sqrt{\frac{c_0 \ln T}{T}} \label{cone} \\ 
  c_2 &= 1 - \mathrm{e}^{-\nu} + \sqrt{\frac{c_0 \ln T}{T}} , \label{ctwo}
\end{align}
then,} recalling from \eqref{eq:T_choice} that $T = \Theta(K \ln N)$, the final two terms in \eqref{eq:term2} vanish at rate $O(T^{-1})$ as $N \to \infty$. Thus, overall, \eqref{eq:term1} and \eqref{eq:term2} simplify to
\begin{gather}
    K \pr(A_i) \geq K c_1^L - \rev{O\Big(\frac{1}{T}\Big)} \label{eq:term1a} \\
    \binom{K}{2} \pr(A_i \cap A_j) \leq \frac{K^2}{2(1-\epsilon_1)}\, \bigg(c_2 + \frac{\nu}{K}\bigg)^{2L} + \rev{O\Big(\frac{1}{T}\Big)}. \label{eq:term1b}
\end{gather}
Combining these, we find that \eqref{eq:SSSfailure_rep} yields
\begin{align}
    \err{SSS} 
        &\geq K c_1^L \left( 1 - \frac{K (c_2 + \nu/K)^{2L}}{ \vphantom{\hat1} 2c_1^L (1-\epsilon_1) } \right) - \rev{O\Big(\frac{1}{T}\Big)} \label{eq:SSSfailure2} \\
        &\rev{= K c_1^{\nu T/K} \left( 1 - \frac{K (c_2 + \nu/K)^{2{\nu T/K}}}{ \vphantom{\hat1} 2c_1^{\nu T/K} (1-\epsilon_1)} \right) - O\Big(\frac{1}{T}\Big)}, \label{eq:SSSfailure2a}
\end{align}
\rev{where we have used the fact that $L = \nu T/K$.}

\proofsubbit{Bounding the right-hand side of \eqref{eq:SSSfailure2a}}
\rev{With the lower bound \eqref{eq:SSSfailure2a} on the error probability in place, the completion of the proof amounts to two simple but tedious steps:
\begin{enumerate}
    \item Equate the large bracketed term with $1 - \frac{1}{2(1-\epsilon_1)} \approx \frac{1}{2}$, and show that solving for $T$ yields an expression of the form \eqref{eq:T_choice};
    \item Show that under any choice of $T$ of the form \eqref{eq:T_choice}, the remaining term $K c_1^{\nu T/K}$ approaches $1$ as $N \to \infty$.
\end{enumerate}
These steps are summarised in the following lemma, whose proof is relegated to Appendix \ref{sec:pf_sss_probs}.}

\begin{lemma} \label{lem:sss_set_vars}
    \rev{Choosing $c_1$ and $c_2$ as in \eqref{cone} and \eqref{ctwo}, there exists a choice of $T$ satisfying \eqref{eq:T_choice} for which \eqref{eq:SSSfailure2a} can be weakened to
        \begin{equation}
        \err{SSS} \geq (1-o(1)) \bigg( 1 - \frac{1}{2(1-\epsilon_1)} \bigg) - O\Big(\frac{1}{T}\Big). \label{eq:SSSfailure3a}
        \end{equation} }
\end{lemma}
\rev{We conclude the proof of Theorem \ref{thm:previous2} by noting that the right-hand side can be made arbitrarily close to $\frac{1}{2}$ for sufficiently large $K$ and $T$, since $\epsilon_1$ can be chosen arbitrarily small.}

\end{IEEEproof}

\proofbit{Conditional distributions of $M_i$ and $G$} 
 
Recall that $M_i$ denotes the number of tests containing defective item $i$ but no other defective items.  The following lemma gives the distribution of this quantity conditioned on $\WKi$, the number of tests covered by those other defectives. It is written in terms of the following definition: For any integers $n,k$ the \emph{Stirling number of the second kind} is given by
 \begin{equation} \label{eq:stirdef}
 \stir{n}{k} := \frac{1}{k!} \sum_{j=0}^k (-1)^{k-j}  \binom{k}{j} j^n,
 \end{equation}
and equals the number of partitions of a set of size $n$ into $k$ nonempty subsets (see for example \cite[eq. (8)]{lieblc}).

\begin{lemma} \label{prop:wdef} \mbox{ }
\begin{enumerate}
\item We can write the conditional distribution of 
$M_i \mid \WKi$ explicitly as 
\begin{equation} \begin{split}
 \pr &\big( M_i = j \mid  \WKi = w \big) \\
 &\qquad\qquad\qquad =  \frac{ (T-w)_{(j)}}{T^L} \sum_{s=0}^{L-j} \binom{L}{s} \stir{ L -s}{j} w^s, \end{split} \label{eq:mkdist} \end{equation}
 where $(n)_{(j)} := (n)!/(n-j)!$ denotes the falling factorial.
\item \rev{For fixed $L$ and $w$, there exists an explicit value $C = C(L,w) := \exp( L^2/4w)$, independent of $j$,} such that
$$ \pr \big( M_i = j \mid  \WKi = w \big)  \leq C \binom{L}{j} \left( 1- \frac{w}{T} \right)^j \left( \frac{w}{T} \right)^{L-j}.$$
That is, the probability is upper bounded by a multiple of the $\bin(L,1-w/T)$ mass function.
\end{enumerate}
\end{lemma} 

\begin{IEEEproof}
See Appendix \ref{sec:mkdist}. \end{IEEEproof}

Next, we observe that  %(see Figure \ref{fig:itemrvs})
 we can write $\WK = \WKi + M_i$.
Recall that $G$ is the number of  non-defectives masked by the defective set $\K$. Since an item is only counted in $G$ if each of the tests appearing in the corresponding column are in the set of size $\WK$, we have the following.

\begin{lemma} \label{lem:gdef}
Conditional on $\WK = x$, we have
$$ G \mid  \big\{ \WK = x \big\} \sim \bin\big(N - K,  (x/T)^L\big).$$
\end{lemma}

\proofbit{Proof of COMP maximum achievable rate}
We can now prove Theorem \ref{thm:previous1} using the above results.

\begin{IEEEproof}[Proof of Theorem \ref{thm:previous1}]
We start with the achievability part, for which we set $\nu = \ln 2$. We consider the regime $T = \gammaCOMP K \ln N$, where $\gammaCOMP = (1+\epsilon)/(\ln 2)^2$.
As mentioned in \eqref{eq:ddsuc}, COMP succeeds if and only if $G = 0$. Using Lemma \ref{lem:gdef}, we know
that 
\begin{equation} \label{compeqn}
\pr( G = 0 \mid  \WK = x) = \left( 1 - \left( \frac{x}{T} \right)^L \right)^{N-K},
\end{equation}
which is a decreasing function in $x$. Hence, given $\delta$, for all $x \leq (1/2 + \delta) T$, we have
\begin{align*}
\pr \big( G = 0 \mid  \WK = x\big) & \geq  \pr\big(G = 0 \bigmid \WK= (1/2 + \delta) T\big) \\
& = \big( 1 - ( 1/2 + \delta )^L \big)^{N-K}.
\end{align*}
Next, using the fact that
  \[L = \frac{T \ln 2}{K} = \gammaCOMP \ln 2\, \ln N = (1+ \epsilon) \frac{1}{\ln 2}\ln N , \]
we find that for any $\epsilon$, we can choose
$\delta$ sufficiently small that $(\frac12 + \delta)^L \leq N^{-(1+\epsilon/2)}$, and hence
$$ \pr \big( G = 0 \mid  \WK = x\big) \geq \big( 1 - N^{-(1 + \epsilon/2)} \big)^{N-K}.$$
We deduce that the success probability $\suc{COMP}$ is lower bounded as follows:
\begin{align*}
    \suc{COMP} &=  \sum_x  \pr \big( \WK = x \big) \pr\big( G = 0 \mid  \WK = x\big) \\
& \geq   \sum_{x \leq (1/2 + \delta)T}  \pr \big( \WK = x \big) \big( 1 - N^{-(1 + \epsilon/2)} \big)^{N-K} \\
& =  \big( 1 - N^{-(1 + \epsilon/2)} \big)^{N-K} \left( 1 - \pr  \big( \WK \geq (1/2 + \delta) T \big) \right),
\end{align*} 
which is seen to tend to $1$ by 
taking $\alpha = \ln 2$ in  Lemma \ref{lem:mcdiarmid} (since we
collect a total of $KL = T \ln 2$ coupons).

The converse proceeds similarly, except that we need to consider a general choice of the parameter $\nu$. By a similar argument to the one above, we deduce that the success probability $\suc{COMP}$ is given by
\begin{align*}
  \sum_x & \pr \big( \WK = x \big) \pr\big( G = 0 \mid  \WK = x\big) \\
& \leq  \pr \big( \WK \leq (1 - \ee^{-\nu} - \delta)T \big) \\&\qquad + \sum_{x \geq (1 - \ee^{-\nu} - \delta)T}  \pr \big( \WK = x \big) \pr\big( G = 0 \mid  \WK = x\big) \\
& \leq  \delta + \pr\big( G = 0 \mid  \WK = (1 - \ee^{-\nu} - \delta)T\big)
\end{align*}
for $N$ sufficiently large, where we have used Lemma \ref{lem:mcdiarmid}.

Using \eqref{compeqn} again, but with $L = \nu T/K$, we have
\begin{equation}
\pr\big( G = 0 \mid  \WK = (1 - \ee^{-\nu} - \delta)T\big)
= \big( 1 - ( 1 - \ee^{-\nu} - \delta )^{\nu T/K} \big)^{N-K}
\end{equation}
Since $(1 - \ee^{-\nu})^\nu$ is minimised at $\nu = \ln 2$, the same is true of the right-hand side when $\delta = 0$.  More generally, we can choose some $\delta'$ (as a function of $\delta$) such that $\delta' \to 0$ as $\delta \to 0$, and continue as follows:
\begin{align*}
&\pr\big( G = 0 \mid  \WK = (1 - \ee^{-\nu} - \delta)T\big) \\
	&\qquad\qquad\leq \left( 1 - \left( \tfrac12 - \delta' \right)^{(\ln 2 + \delta') T/K} \right)^{N-K}.
	% &\qquad\qquad\leq 1 - (N-K)\left( \tfrac12 - \delta' \right)^{(\ln 2 + \delta') T/K} \nonumber \\ &\qquad\qquad\qquad \times\left(1 - \frac{N-K}{2} \left( \tfrac12 - \delta' \right)^{(\ln 2 + \delta') T/K}\right).
\end{align*}
Since
  \[ \lim_{N \to \infty} \left(1 - \frac{c}{N-K}\right)^{N-K} = \mathrm e^{-c} \]
for any $c > 0$, we find that the error probability is upper bounded by $\ee^{-c} (1+o(1))$ when $(N-K)\big( \tfrac12 - \delta' \big)^{(\ln 2 + \delta') T/K} \ge c$.  Taking the logarithm and noting that $c$ and $\delta'$ can be arbitrarily small, we find that the success probability vanishes when
  \[ \ln 2 \, \frac{(\ln 2)T}{K \ln N}  \leq 1 - \epsilon , \]
which is precisely when $T \leq (1 - \epsilon) T^{\mathrm{COMP}}$.
\end{IEEEproof}

\proofbit{Conditional distribution of $L_i$} Recalling that $L_i$ denotes the number of tests containing defective item $i$ and no other ``possible defective'' (item from $\PD$), we have the following.  

\begin{lemma} \label{lem:plkzero} For any $g$, $w$, $j$, we have
$$ \pr\big( L_i = 0 \bigmid G= g,  \WKi = w, M_i = j\big) = \phi_j \left( \frac{1}{w+j}\,, g L \right),$$
where 
\begin{equation} \label{eq:phidef}
\phi_j(s,V) = \sum_{\ell=0}^j (-1)^\ell \binom{j}{\ell} \left( 1 - \ell s \right)^V.
\end{equation}
\end{lemma}
\begin{IEEEproof} See Appendix \ref{sec:mkdist}. \end{IEEEproof}

%%%%%%%%%%
% Big displayed equation out of place,
%to try and force it to the top of the desired page
%%%%%%%%%%
% \begin{figure*} \label{displayeqn}
%\setcounter{MYtempeqncnt}{\value{equation}}
%\setcounter{equation}{15}

% [Jon: For the single-column submission I've moved it back to the main text]

%\setcounter{equation}{\value{MYtempeqncnt}}
% \hrulefill
% \end{figure*}

Note that the function $\phi_j(s,V)$ also appeared in \cite{johnson33}; however, our analysis here requires using it very differently.  We make use of the following properties, the proofs of which are deferred to Appendix \ref{sec:phipropproofs}.

\begin{lemma} \label{lem:phiupdown}
For all values of $j$, $s$ and $V$, the function $\phi_j(s,V)$ introduced in 
\eqref{eq:phidef} has the properties that:
\begin{itemize}
\item $\phi_j(s,V)$ is increasing in $s$,
\item $\phi_j(s,V)$ is increasing in $V$,
\item $\phi_j(s,V)$ is decreasing in $j$.
\end{itemize}
\end{lemma}

\begin{lemma} \label{prop:phibd} If $s V j \leq 2$,  then
$$  \phi_j(s, V) \leq \frac{V! s^j  }{(V-j)!} 
\leq \exp \left( j \ln( V s) \right). $$
\end{lemma}

\proofbit{Proof of the DD achievable rate}
 We  put the above results together to
 prove Theorem \ref{thm:ddrate}, giving a lower bound on the achievable rate of the DD algorithm. The key  is to express the success probability $\suc{DD}$ in terms of  an expectation of the function 
$\phi$, and to  show that this expectation is concentrated in a regime where $\phi$ takes favourable values.

\begin{IEEEproof}[Proof of Theorem \ref{thm:ddrate}]
We consider the regime where
 $T = \gammaDD m K \ln N$, with $\gammaDD = (1+\epsilon)/(\ln 2)^2$ and $m = \max\{\theta,1-\theta\}$.  In addition, we choose the parameter $\nu = \ln 2$.
As a result, $L = \nu T/K$ satisfies the following:
  \[ L \ln 2 = \frac{T (\ln 2)^2}{K} = m(1+\epsilon) \ln N . \]

As in \cite{johnson33}, writing $\suc{DD}$ for the success probability of DD and
applying the union bound to \eqref{eq:ddsuc} we know that
\begin{equation} \label{eq:succprob}
 \suc{DD} = 1 - \pr \left( \bigcup_{i \in \K}  \{ L_i  =  0 \} \right)
\geq 1 - \sum_{i \in \K} \pr(  L_i = 0), \end{equation}
so that $\suc{DD}$ will tend to 1 (as required)  if, for a particular defective item $i \in \K$,
\begin{equation} \label{eq:required}
   K \pr( L_i = 0) \rightarrow 0,
\end{equation}
since symmetry means that $\pr(L_i = 0)$ is equal for each $i \in \K$. The stated value for the rate then follows upon substituting the 
choice $T = \gammaDD m K \ln N$ and \eqref{eq:binasymp} into \eqref{eq:ratedef}.

\begin{figure*}[!t]
\begin{align}  \pr(L_i = 0 )
& =  \sum_{w,j,g} \pr\big(\WKi = w, M_i = j, G = g\big) \times \big( \II[  A \cap B  ] + \II \left[ (A \cap B )^\com \right]  \big) \phi_j
\left( \frac{1}{w+j} ,gL \right) \nonumber \\
& \leq   \sum_{w \in [w_-,w_+]} \pr\big(\WKi = w\big)  \sum_{j=0}^L \pr\big(M_i = j \bigmid \WKi = w\big)  \phi_{j}(1/w_-, g^*L) + \pr \big( (A \cap B)^\com \big) \label{eq:val1} \\
& =   \sum_{w \in [w_-,w_+]} \pr\big(\WKi = w\big) \sum_{j=0}^L \pr\big(M_i = j \bigmid \WKi = w\big)  \phi_{j}(1/w_-, g^*L)   % \notag \\ &\qquad\qquad {}
+  \pr(A^\com) + \pr(A \cap B^\com) 
\nonumber \\
\begin{split}
& \leq   \sum_{w \in [w_-,w_+]} \pr\big(\WKi =w\big) \sum_{j=0}^L \pr\big(M_i = j \bigmid \WKi = w\big)  \phi_{j}(1/w_-, g^*L)
      \\
&   \qquad {}
+ \pr\big(\WKi\notin [w_-,w_+]\big) + \pr\big( G > g^* \bigmid \WKi\in [w_-,w_+] \big), \label{eq:bigsum} \end{split}
\end{align}
\hrulefill
\vspace*{4pt}
\end{figure*}

In order to characterise $\pr( L_i = 0)$, we define $A = \{ w_- \leq \WKi\leq w_+ \}$ and $B = \{ G \leq g^* \}$, for some $w_-$, $w_+$ and $g^*$ to be chosen shortly.
Using Lemma \ref{lem:plkzero}, we have
the terms at in the large displayed equations at the top of the following %%%% Check "following"/"previous" is still correct
page, where:
\begin{itemize}
	\item \eqref{eq:val1} follows because, by Lemma \ref{lem:phiupdown},  on the event $\{ A \cap B \}$ the bound $\phi_j(1/w, g L) \leq \phi_{j}(1/w_-, g^*L)$ holds, and everywhere else  $\phi \leq 1$ (since $\phi$ represents a probability);
   	\item \eqref{eq:bigsum} follows since
    $\pr(A \cap B^\com) = \pr(B^\com \mid A) \pr(A) \leq \pr(B^\com  \mid A)$.
\end{itemize}

We consider the terms of \eqref{eq:bigsum} separately, taking $w_- = T(1-\delta)/2$, $w_+ = T(1+\delta)/2$, and $g^* = N (1/2 + \delta)^L$, where $\delta = (\epsilon \ln 2) / 4(1+\epsilon)$.

 The first term of \eqref{eq:bigsum}
can be bounded as follows.  Combining $L = (\ln 2) T/K$ and $w_- = T(1-\delta)/2$ gives $L/w_- = 2 \ln 2/(K (1-\delta))$, and recalling that $g^* = N (1/2 + \delta)^L$ and $m = \max(\theta, 1-\theta)$, it follows that
\begin{align}
\beta & := \ln \left( \frac{g^* L}{w_-} \right) \notag \\
& =   (1-\theta) \ln N + L \ln(1/2 + \delta) + \ln \left( \frac{2 \ln 2}{1-\delta}
\right)  \notag \\
& \leq   m \left( 1 + (1+\epsilon) \Big(-1 + \frac{2 \delta}{\ln 2} \Big)  \right) \ln N
+ \ln \left( \frac{2 \ln 2}{1-\delta} \right)  \notag \\
& \leq   m \left( - \frac{\epsilon}{2}  \right)  \ln N
+ \ln \left( \frac{2 \ln 2}{1-\delta} \right), \label{eq:betabd}
\end{align}
where the second line follows by combining $L = m(1+\epsilon)\ln N /\ln 2$ and $\ln(1/2 + \delta) \le -\ln 2 + 2\delta$, and the third line follows since $1 + (1+\epsilon) \left(-1 + 2 \delta/\ln 2 \right) \le -\epsilon / 2$ under the above choice $\delta = (\epsilon \ln 2) / 4(1+\epsilon)$.
We claim that \eqref{eq:betabd} implies $j g^* L/w_- \leq 2$ for all $j \le L$.  Indeed, we have $T = \Theta(K \log N)$ and $L = \Theta(T/K)$, so that $L = \Theta(\log N)$, whereas \eqref{eq:betabd} implies that $g^* L/w_-$ decays to zero strictly faster than $1/\log N$.  This implies that
\begin{equation} \label{phibd}
 \phi_{j}(1/w_-, g^*L)  \leq  \exp \left( j \ln \Big( \frac{g^* L}{w_-} \Big) \right) = \ee^{j\beta} ,
\end{equation}
since the conditions of Lemma \ref{prop:phibd} are satisfied under these arguments.
Writing $\phi(j) = \phi_j(1/w_-, g^* L)$ (which is decreasing in $j$ by the third part f Lemma \ref{lem:phiupdown}), we can
bound $K$ times the inner sum of \eqref{eq:bigsum} as follows:
\begin{align}
 K   \sum_{j=0}^L & \,\pr\big(M_k = j \; \big|\; \WKi = w\big)\, \phi(j)   \notag\\
 &\leq K C(L,w) \sum_{j=0}^L  \pr\big( \bin(L, 1-w/T) = j \big) \phi(j)  \label{eq:Mk_step1} \\
\begin{split} & = K C(L,w) \sum_{j=0}^L  \pr\big( \bin(L, 1-w_+/T) = j \big)  \\
& \qquad \times \rev{\left(
  \frac{\pr( \bin(L, 1-w/T) = j ) }{\pr( \bin(L, 1-w_+/T) = j )} \right)} \, \phi(j)  \end{split}\label{eq:cheby} \\
&\leq K C(L,w) \sum_{j=0}^L  \pr\big( \bin(L, 1-w_+/T) = j\big) \phi(j) \label{eq:Mk_step3} \\
& \leq K C(L,w_-) \left( \frac{w_+}{T} + \frac{T-w_+}{T} \, \eb 
\right)^L \label{eq:Mk_step4} 
\end{align}
Here:
\begin{itemize}
	\item \eqref{eq:Mk_step1} follows from the second part of Lemma \ref{prop:wdef}.
    \item \rev{We deduce \eqref{eq:Mk_step3} using the following argument:} The bracketed term in \eqref{eq:cheby} is easily verified to be increasing in $j$ by substituting the Binomial mass function and noting $1-w/T \ge 1-w_+/T$, and we already know from Lemma \ref{lem:phiupdown} that $\phi(j)$ is decreasing.  Hence, \eqref{eq:cheby} is the expectation of the product of an increasing and decreasing function, and so by `Chebyshev's other inequality' \cite[eq.~(1.7)]{kingman}, it is bounded above by the product of the expectations of those functions.\footnote{In fact, \cite[eq.~(1.7)]{kingman} concerns $\ep[f(X)g(X)]$ for two increasing functions, but we can transform this to $\ep[f(X)h(X)]$ for decreasing $h$ by simply defining $h(x) = L - g(x)$.}
    \item \eqref{eq:Mk_step4} follows by upper bounding $C(L,w) \le C(L,w_-)$ and $\phi(j) \le \ee^{j\beta}$ from \eqref{phibd}, and then evaluating the sum exactly.
\end{itemize}
We can simplify \eqref{eq:Mk_step4} using the following: 
\begin{align}
&  \!\!\!\! \rev{K C(L,w_-)} \left( \frac{w_+}{T} + \frac{T-w_+}{T} \,\eb 
\right)^L \notag \\
& \quad =  \frac{K \rev{C(L,w_-)}}{2^L} \big(1 + \delta + \eb(1-\delta) \big)^L \label{eq:Mk_step5} \\
& \quad \leq   \,\rev{C(L,w_-)} \cdot c \exp\left( -\rev{m} \epsilon \ln N \right)  \exp \big( L( \delta + \eb(1-\delta))\big)  \label{eq:Mk_step6} \\
& \quad \leq   \,\rev{C(L,w_-)} \cdot c \exp \left( \Big( \!-\rev{m} \epsilon + \frac{ m(1+\epsilon)}{\ln 2}
\big( \delta + \eb(1-\delta)\big) \Big)  \ln N \right).
\label{eq:boundinnersum}
\end{align}
Here:
\begin{itemize}
    \item \eqref{eq:Mk_step5} follows by substituting $w_+ = T(1+\delta)/2$.
    \item \eqref{eq:Mk_step6} follows from $1+\zeta \le \ee^{\zeta}$, along with the fact that 
      \[ \frac{K}{2^L} \rev{\leq }
      c \exp  \big( \big(m - m(1+\epsilon)\big) \ln N \big)
      \leq c \exp(- \rev{m} \epsilon \ln N) \]
    by $L = m(1+\epsilon)\ln N / \ln 2$ and by $K = \Theta(N^{\theta})$ giving $K \le cN^{\theta} \rev{\le c N^m}$ for some $c = \Theta(1)$.
    \item \eqref{eq:boundinnersum} follows by again using $L = m(1+\epsilon)\ln N / \ln 2$.
\end{itemize}
We conclude that \eqref{eq:boundinnersum} acts as an upper bound on  $K$ times the first term of \eqref{eq:bigsum}.  Overall \eqref{eq:boundinnersum} tends to zero for $\delta$ sufficiently small, since $C(L,w_-) = \exp( L^2/4w_-)$ tends to 1 in this regime.

The second term of \eqref{eq:bigsum} decays to zero exponentially fast in $T$ by Lemma \ref{lem:mcdiarmid}.
More precisely, we make $(K-1) L$ draws with replacement, so that $\alpha = (K-1) \ln  2/K \rightarrow \ln 2$, meaning that
we can take $\epsilon = \delta/3$ in Lemma \ref{lem:mcdiarmid} to obtain 
\begin{align*}
 \limsup_{N \rightarrow \infty}  \,K \,&\pr\big(\WKi\notin (w_-,w_+)\big) \\
& \leq   \limsup_{N \rightarrow \infty}  K\, 
\pr \left( \,\big| \WKi - (1-\ee^{-\alpha}) \,T \,\big| \geq \epsilon T \right)  \\
& \leq  2 \limsup_{N \rightarrow \infty} K \exp \left( \!-\frac{\epsilon^2 T}{\alpha} \right) \\
& =  2c \limsup_{N \rightarrow \infty} \exp \left( \ln N \Big( \theta - \frac{\epsilon^2  \gammaDD m K}{\alpha} \Big) \right),
\end{align*}
since $T = \gammaDD mK \ln N$ and $K = \Theta(N^{\theta})$ (and hence $K \le cN^{\theta}$ for some $c = \Theta(1)$).  We conclude that this term tends to zero, since the exponent behaves as $-K \ln N$.

To control the third term in \eqref{eq:bigsum}, observe that if $\WKi\leq w_+$, then 
\[\frac{\WK}{T} \leq \frac{1+\delta}{2} + \frac{L}{T}
= \frac{1+\delta}{2} + \frac{\ln 2}{K} \leq \frac12 + \frac{3 \delta}{4} , \]
where the first inequality holds since $\WK \le \WKi + L$ and $w_+ = T(1+\delta)/2$, the equality holds since $L = \ln 2 T/K$, and the final inequality holds for $K$ sufficiently large.  Hence, and defining $p = (1/2 + 3 \delta/4)^L$, Lemma \ref{lem:gdef} gives 
\begin{align}
 \pr & \big( G > g^*  \mid \WKi\in (w_-,w_+)\big) \notag \\
 &\qquad \leq  \pr\big( \bin(N, p) > g^*\big) \notag \\
 &\qquad \leq   \exp \left( - \frac{ (g^*)^2}{2( Np + g^*/3 )} \right) \label{eq:p_G_cond3}\\
 &\qquad= \exp \left( - N \frac{ (1/2 + \delta)^{2L} }{ 2\big( (1/2 + 3\delta/4)^{L} + (1/2 + \delta)^{L}/3 \big) } \right) \label{eq:p_G_cond4}  \\
 &\qquad= \exp \left( - N \frac{ (1/2 + \delta)^{L} }{ 2(1/3 + o(1)) } \right), \label{eq:final_term_decay}
\end{align}
where \eqref{eq:p_G_cond3} follows from Bernstein's inequality \cite[eq.~(2.10)]{boucheron}, \eqref{eq:p_G_cond4} follows from $p = (1/2 + 3 \delta/4)^L$ and $g^* = N(1/2 + \delta)^L$, and \eqref{eq:final_term_decay} follows since the ratio of $(1/2 + 3 \delta/4)^L$ to $(1/2 + \delta)^L$ tends to zero as $N \to \infty$ (and hence $L \to \infty$, since $L = \Theta(\log N)$).  

Finally, since $L = m(1+\epsilon)\log N / \log 2$, we find that $(1/2 + \delta)^L$ behaves as $N^{\rev{-c}}$ for some $c$ that can be made arbitrarily close to $m = \max\{\theta,1-\theta\}$ by choosing $\delta$ and $\epsilon$ sufficiently small.  By definition, $m < 1$, and the bound in \eqref{eq:final_term_decay} is exponential in $N^{1-c}$, meaning that it vanishes even when multiplied by $K$.
\end{IEEEproof}

\section{Conclusions and open questions}

We have studied nonadaptive group testing with \constant\ column weight designs. We have seen that:
\begin{itemize}
\item The very simple COMP algorithms requires $23.4\%$ fewer tests with a \constant\ column weight design than with a Bernoulli design, performing even better than optimal algorithms with Bernoulli designs for $\theta > 0.766$.
\item Using a \constant\ column weight design, the practical DD algorithm
again uses $23.4\%$ fewer tests than with Bernoulli designs,  outperforms any possible algorithm for Bernoulli designs for $\theta > 0.434$, and beats the best-known theoretical guarantees of existing practical algorithms for all $\theta \in (0,1)$.
\item An upper bound on performance of \constant\ column weight designs shows that DD is optimal for this design when $\theta \geq 1/2$.
\item Numerical experiments demonstrated a notable improvement over Bernoulli designs in both sparse and dense regimes.
\end{itemize}

We briefly mention \rev{some} interesting open problems connected with this paper, which we hope to address in future work:
\begin{enumerate}
\item It remains open to determine the maximum achievable rate of constant or near-constant column weight designs for $\theta \leq 1/2$, in the spirit of Theorem \ref{thm:berncap}. We conjecture that the value is \eqref{eq:conv} (i.e., Theorem \ref{thm:previous2} is sharp), and is achievable by the
maximum likelihood algorithm (as well as the SSS algorithm of \cite{johnson33}). This is the value suggested by a non-rigorous result of M\'ezard, Tarzia and Toninelli \cite{mezard}.
\item It is an important open problem to determine whether `practical' algorithms can improve on the performance of DD. For example, the SCOMP algorithm of \cite{johnson33} and approaches based on linear programming both have a rate at least as large as DD \cite{aldridge-dd}. However, we do not know the {\em best possible} rate of DD for $\theta < 1/2$, nor how to determine whether these algorithms or others can have a higher rate than DD.%\footnote{Note that the upper bound for DD with Bernoulli designs in \cite{aldridge-dd} requires independent tests, so does not apply to constant column-weight designs here, as we erroneously claimed in a preprint of this paper.}
\item It remains of great interest to determine whether a rate of $1$ can be achieved for values of $\theta$ beyond $1/3$ using constant or near-constant column weights or some other design. The conjecture above would imply that \constant\ column-weight designs achieve rate 1 for $\theta < 0.409$. More generally, it is an open problem as to whether there exists an `adaptivity gap', i.e., a choice of $\theta < 1$ such that any nonadaptive design must have rate less than $1$, despite the rate of $1$ being achievable  with adaptive testing.
% \item A design of potential interest is that with both (near-)constant column weights (tests-per-item) and also (near-)constant row weights (items-per-test). On one hand, the nonrigorous work of \cite{mezard} suggests that for $\theta < 1$ there may be no gain over constant column weight designs considered here. On the other hand, in light of Wadayama's work \cite{wadayama} on constant row-and-column designs in the regime where $K$ grows linearly with $N$, and the performance of LDPC codes, this does seem like a natural place to look for improvements. Either way, the independence of the columns of $\mat{X}$ plays a significant role in the proofs of this paper, so new arguments would be required to investigate this.
\end{enumerate}

% PREVIOUS VERSION OF LAST BULLET POINT

%\item It remains of great interest to know what the optimal design is, and whether a capacity of $1$ can be achieved for all values of $\theta$ using some other design. In the light of Wadayama's work \cite{wadayama}, and the performance of LDPC codes, it natural to conjecture that in some circumstances better performance than for constant column weights will be obtained by constant row-and-column weight designs. However, since the independence of columns of $\mat{X}$ plays a significant role in the proofs of this paper, new arguments would be required to prove this.

\appendices

\section{Properties of the distribution of $M_i$} \label{sec:mkdist}

\subsection{Proof of Lemma \ref{prop:wdef}}

\begin{IEEEproof}[Proof of Lemma \ref{prop:wdef}] 
We prove the first part of the lemma directly. Suppose that we pick $L$ coupons from a population of $T$ coupons, $w$ of which were previously chosen.
Clearly, the probability of the event that exactly $s$ of the coupons picked
were previously chosen is $\pr( \bin(L, w/T) = s)$.

Conditioning on this event, we calculate the probability that we pick $L-s$ coupons out of a population of $T-w$ coupons  and obtain exactly $j$ distinct new coupons. Clearly
we require $L-s \geq j$, or $s \leq L-j$. By a standard counting argument,
we can choose these $j$ coupons $\binom{T-w}{j}$ ways, then $\smallstir{L-s}{j}$ ways of placing the $L-s$ coupons
into $j$ unlabelled bins such that none of them are empty (see \cite[p.~204]{lieblc}), and finally $j!$ different labellings of the bins.
Moreover, overall there are $(T-w)^{L-s}$ assignments of the coupons.

Putting this all together and recalling the definition
  \[ (T-w)_{(j)} = \frac{(T-w)!}{(T-w-j)!} = \binom{T-w}{j} j! , \]
we have
\begin{align*}
\pr &\big( M_{i} = j \bigmid  \WKi = w \big)  \\
  & = \sum_{s=0}^{L-j} \binom{L}{s} \left( \frac{w}{T} \right)^s \left( 1  - \frac{w}{T} \right)^{L-s} 
\binom{T-w}{j}  \stir{L-s}{j}  j!  \,\frac{ 1 }{(T-w)^{L-s}}   \\ 
  & = \sum_{s=0}^{L-j} \binom{L}{s} \frac{w^s}{T^L} \, 
 (T-w)_{(j)}   \stir{L-s}{j},
\end{align*}
as required.

We now prove the second part of the lemma. Relabelling $t = L-j-s \geq 0$ and
using the fact that $\smallstir{t+j}{j} \leq \binom{t+j}{j} j^t$ (see \cite{rennie}),
we obtain that the inner sum of \eqref{eq:mkdist} is:
\begin{align*}
 \sum_{t=0}^{L-j}   \binom{L}{L-j-t} &\stir{ t+j}{j} w^{L-j-t}  \\
 & \leq  w^{L-j}  \sum_{t=0}^{L-j} \binom{L}{L-j-t} \binom{t+j}{j} \Big( \frac{j}{w} \Big)^t \\
 & =  w^{L-j} \binom{L}{j} \sum_{t=0}^{L-j} \binom{L-j}{t} \Big( \frac{j}{w} \Big)^t \\
& =  w^{L-j} \binom{L}{j} \Big( 1 + \frac{j}{w} \Big)^{L-j} \\
& \leq  w^{L-j} \binom{L}{j} C,
\end{align*}
where the third line follows by explicitly evaluating the summation, and the final line holds with $C = \exp(L^2/4w)$ since
\[ \left(1 + \frac jw \right)^{L-j} \leq \exp\left( \frac{j (L-j)}{w}\right) \le \exp\left(\frac{L^2}{4w}\right) . \]
This allows us to deduce that the whole of \eqref{eq:mkdist} satisfies
\begin{align*}
\pr\big(M_i = j  \bigmid  \WKi= w \big) 
& \leq  C \frac{ (T-w)_{(j)}}{T^L}  \, w^{L-j} \binom{L}{j} \\
& \leq C \frac{ (T-w)^j}{T^L}  \, w^{L-j} \binom{L}{j}, 
\end{align*}
as required.
\end{IEEEproof}

\subsection{Proof of Lemma \ref{lem:plkzero}}

\begin{IEEEproof}[Proof of Lemma \ref{lem:plkzero}]
The case $j = 0$ is trivial, so we assume here that $j \ge 1$.

We have conditioned on three events: on $\WKi = w$, meaning there are $w$ tests containing one or more item from $\K \setminus {i}$; on event $M_i = j$, meaning there are $j$ tests that contain item $i$ and no member of $\K \setminus {i}$; and on $G=g$, meaning there are $g$ items labelled as possibly defective but not in $\K$.

By relabelling, without loss of generality, we can assume that tests $1, \ldots, j$ are the ones that contain defective
item $i$ and no other defective item.
We write $A_s$ for the event that test $s$ does not have any element  of
$\PD \setminus \K$ in it.

If an item is in $\PD \setminus \K$, then the tests that it appears in are chosen uniformly among those which already contain a defective.
Hence, for any set $S \subseteq \{1,\dotsc,j\}$ of size $\ell$, we have $ \pr ( \bigcap_{r \in S} A_r) = ( 1 - \ell/(w+j) )^{g L}$. This is because we require that the $L$ coupons of each of $g$ items in $\PD \setminus \K$ take values in the set of positive tests ($\WKi + M_i = w+j$ in total), but avoid the specified $\ell$ tests.  Thus,
\begin{align} 
\pr\big( L_i = 0  \bigmid G= g,  & \,\WKi = w, M_i = j\big) \nonumber \\
& =  \pr \left( \bigcap_{s=1}^j A_s^\com \right) \nonumber  \\
& =  1 - \pr \left( \bigcup_{s=1}^j A_s \right) \nonumber \\
& =   \sum_{\ell=0}^j  (-1)^\ell \sum_{\substack{S \subseteq \{1,\dotsc,j\} \\ |S| = \ell}} \pr \left( \bigcap_{j \in S} A_j \right)  \nonumber \\
& =  \sum_{\ell=0}^j (-1)^\ell \binom{j}{\ell} \left( 1 - \frac{\ell}{w+j} \right)^{g L}, \label{eq:phifunc}
\end{align}
and the result follows.
\end{IEEEproof}

\section{Properties of the $\phi$ function} \label{sec:phipropproofs}

\subsection{Proof of Lemma \ref{lem:phiupdown}}

\begin{IEEEproof}[Proof of Lemma \ref{lem:phiupdown}]
We deduce the results using the expression
  \[ \phi_j(s,V) = \sum_{\ell=0}^j (-1)^\ell \binom{j}{\ell} \left( 1 - \ell s \right)^V \]
from \eqref{eq:phidef}.

First we show that $\phi_j(s,V)$ is increasing in $s$.
As in \cite[Lemma 32]{johnson33}, a direct calculation using the fact that $\ell \binom{j}{\ell} = j \binom{j-1}{\ell-1}$ gives
\begin{align}
\frac{\partial}{\partial s} & \phi_j(s,V)  \notag \\
& =  V j \sum_{\ell =1}^j (-1)^{\ell-1} \binom{j-1}{\ell-1} \left( 1 - \ell s \right)^{V-1} \label{eq:firstder} \\
& =  (1-s)^{V-1} V j \sum_{\ell =1}^j (-1)^{\ell-1} \binom{j-1}{\ell-1} \left( 1 -  \frac{(\ell -1) s}{1-s} \right)^{V-1}  \nonumber \\
& =  (1-s)^{V-1} V j \,\phi_{j-1} \left( \frac{s}{1-s}, V-1 \right) \\
 &\geq 0, \nonumber 
\end{align}
where the second line uses the fact that
\[ (1- \ell s) = (1-s) \left( 1- \frac{(\ell -1) s}{1-s} \right) , \]
and the third line above follows by relabelling $\ell' = \ell - 1$. 

Second, we show that $\phi_j(s,V)$ is increasing in $V$.
Again using $\ell \binom{j}{\ell} = j \binom{j-1}{\ell-1}$, we can write
\begin{align*}
\phi_j(s,V) & - \phi_j(s,V-1)  \\
& = 
\sum_{\ell=0}^j (-1)^{\ell} \binom{j}{l} (1- \ell s)^{V-1} \big((1- \ell s) - 1\big) \\ 
& =   s j  \sum_{\ell =1}^j (-1)^{\ell-1} \binom{j-1}{\ell-1} \left( 1 - \ell s \right)^{V-1} \\
& =   \frac{s}{V} \frac{\partial}{\partial s} \phi_j(s,V) \\
&\geq 0,
\end{align*}
where the third line follows from \eqref{eq:firstder}.

Third, we show that $\phi_j(s,V)$ is decreasing in $j$. By expanding $\binom{j}{\ell} = \binom{j-1}{\ell} + \binom{j-1}{\ell-1}$, we can write
\begin{align*}
\phi_j(s,V) & =  \sum_{\ell=0}^j (-1)^\ell \left( \binom{j-1}{\ell} + \binom{j-1}{\ell-1} \right) \left( 1 - \ell s \right)^V \\
& =  \phi_{j-1}(s,V) - \frac{1}{(V+1) j}  \frac{\partial}{\partial s} \phi_j(s,V+1) \\
& \leq  \phi_{j-1}(s,V),
\end{align*}
again using \eqref{eq:firstder}.
\end{IEEEproof} 

\subsection{Proof of Lemma \ref{prop:phibd}}

We now prove Lemma \ref{prop:phibd}, first giving two preliminary lemmas.

\begin{lemma} \label{lem:phipoly}
We can expand $\phi_j(s,V)$ (as defined in \eqref{eq:phidef})
as a polynomial in $s$ of degree $V$ as follows:
\begin{equation}
 \phi_j(s, V) = \frac{\rev{V! s^j}}{(V-j)!} \sum_{u=0}^{V-j} (-1)^u s^u \frac{j!  (V-j)!}{(u+j)! (V-u-j)!} \stir{j+u}{j}, \label{eq:phiexp}
\end{equation}
where we again write $\smallstir{j+u}{j}$ for the Stirling number of the second kind.
\end{lemma}

\begin{IEEEproof}
We can expand
\begin{align} 
 \phi_j(s, V)  & =  \sum_{\ell=0}^j (-1)^\ell \binom{j}{\ell} \left( 1 - \ell s \right)^V \nonumber \\
 & =  \sum_{\ell=0}^j (-1)^\ell \binom{j}{\ell} \sum_{t=0}^V  \binom{V}{t} (-s)^t \ell^t \nonumber \\
 & =  \sum_{t=0}^V  \binom{V}{t}  (-s)^t 
 \sum_{\ell=0}^j (-1)^\ell \binom{j}{\ell} \ell^t  \label{eq:stirbd} \\
 & =  \sum_{t=0}^V  \binom{V}{t}  (-s)^t \stir{t}{j} j!\, (-1)^j  \nonumber \\
 & = \sum_{u=0}^{V-j}  \binom{V}{j+u}  (-s)^{j+u}  \stir{j+u}{j} j! \,(-1)^j  \nonumber,
\end{align}
where the second line can be seen by directly evaluating the summation, the fourth line follows by recognising the inner sum in \eqref{eq:stirbd}
as a multiple of the Stirling number using \eqref{eq:stirdef}, and the last line follows since by relabelling $t = j + u$ and noting that $\smallstir{t}{j}$ is
non-zero only when $t \geq j$.  The result now follows by writing 
\[ \binom{V}{j+u} = \frac{V!}{(V-u-j)!\,(u+j)!} \; \frac{(V-j)!}{(V-j)!} \]
and $(-s)^{j+u} (-1)^j = (-1)^u s^js^u$.
\end{IEEEproof}

We also use the following result from \cite[Theorem 4.4]{neuman}.
\begin{lemma} \label{lem:stirlc} The Stirling numbers of the second kind are log-concave in their first argument, that is  for
any $j,u \in \Z_+$:
$$  \stir{j+u+1}{j}^2 \geq \stir{j+u}{j} \stir{j+u+2}{j}.$$
\end{lemma}

We are now in a position to prove Lemma \ref{prop:phibd}.

\begin{IEEEproof}[Proof of Lemma \ref{prop:phibd}] Using Lemma \ref{lem:phipoly},
we consider $\phi_j(s,V)$ as a sum of the form
$$ 
 \phi_j(s, V) = \frac{s^j V!}{(V-j)!} \sum_{u=0}^{V-j} (-1)^u a_u,$$
where
$$ a_u = s^u \frac{j! \, (V-j)!}{(u+j)! \,(V-u-j)!} \stir{j+u}{j}.$$
By the alternating series test, if $a_u$ is a monotonically decreasing sequence, we can bound
$\sum_{u=0}^{V-j} (-1)^u a_u \leq a_0 = 1$,
and the result follows.
We can verify that $a_u$ is indeed monotonically decreasing by considering the ratio
\begin{equation} \label{eq:ratioas}
\frac{a_{u+1}}{a_u} = s \frac{V-j-u}{j+u+1}  \; \frac{ \smallstir{j+u+1}{j}}{\smallstir{j+u}{j}} .
\end{equation}
The first fraction in \eqref{eq:ratioas} is trivially decreasing in $u$, and the second fraction in \eqref{eq:ratioas} is decreasing in $u$ 
by Lemma \ref{lem:stirlc}. Hence, since the ratio \eqref{eq:ratioas} is decreasing in
$u$, it is sufficient to verify that $a_1/a_0 \leq 1$. Since $\smallstir{j}{j} =1$ and $\smallstir{j+1}{j} = j(j+1)/2$, 
direct substitution in
\eqref{eq:ratioas} gives that $a_1/a_0 = s(V-j) j/2$, so it is
sufficient to assume that $s(V-j) j/2 \leq 1$.
\end{IEEEproof}

\section{\rev{Auxiliary results for the algorithm-independent converse}} \label{sec:pf_sss_probs}

\subsection{\rev{Proof of Lemma \ref{lem:sss_probs}}} 

Fixing the index $i$ of some defective item, we note that conditioned on $\WKi = w$, the event $A_i$ occurs if each  test that item $i$ occurs in is contained in the $w$ `already hit' tests.  Hence, for any $c_1 > 0$, we can write
\begin{align*}
\pr(A_i) &=  \sum_w \pr\big(A_i \bigmid \WKi = w\big) \pr\big( \WKi = w\big) \notag \\
&=  \sum_w \left( \frac{w}{T} \right)^L \pr \big(\WKi = w\big) \notag \\
&\geq  \sum_w \left( \frac{w}{T} \right)^L \pr \big(\WKi = w\big) \II[ w \geq T c_1] 
\notag \\
&\geq  c_1^L \,\pr\big( \WKi \geq T c_1\big),
\end{align*}
which proves \eqref{eq:term1}.

The analysis of the event $A_i \cap A_j$ for $i \ne j$ is more challenging; we show in Section \ref{sec:pf_intersection} below that if $w \ge T c_1$ with the same $c_1 > 0$ as above, then for arbitrarily small $\epsilon_1 > 0$ it holds for sufficiently large $N$ that
%for any $c_2 > 0$ we have 
%\begin{align}
%\binom{K}{2} &  \pr(A_i \cap A_j)\notag \\
%  &=\binom{K}{2} \sum_w \pr\big(A_i \cap A_j \bigmid \WKij = w\big) \, \pr\big( \WKij = w\big) \nonumber \\
%& = \binom{K}{2} \sum_w \left( \frac{w}{T} \right)^{2L} \pr \big(\WKij = w\big) \nonumber \\
%& \leq \binom{K}{2} \sum_w \left( \frac{w}{T} \right)^{2L} \pr \big(\WKij = w\big) \,\II[ w \leq T c_2]  \nonumber \\
%& \qquad {}+ \binom{K}{2} \pr\big( \WKij \geq T c_2\big) \nonumber \\
%& \leq \frac{K^2}{2}\, c_2^{2L} \,\pr\big( \WKij \leq T c_2\big) + 
%\binom{K}{2} \pr\big( \WKij \geq T c_2\big). \label{eq:term2}
%\end{align} 
\begin{equation}   
\pr\big(A_i \cap A_j \mid \WKij = w\big) \leq \left(\frac{w+L}{T}\right)^{2L} \frac{1}{1-\epsilon_1}. \label{eq:intersection}
\end{equation}
Hence, for any $c_1,c_2 > 0$, we have
\begin{align*}
\binom{K}{2} &  \pr(A_i \cap A_j)\notag \\
&=\binom{K}{2} \sum_w \pr\big(A_i \cap A_j \bigmid \WKij = w\big) \, \pr\big( \WKij = w\big) \nonumber \\
% & \le \frac{1}{1-\epsilon_1} \binom{K}{2} \sum_w \left( \frac{w+L}{T} \right)^{2L} \pr \big(\WKij = w\big) \nonumber \\
& \leq \frac{1}{1-\epsilon_1}\binom{K}{2} \sum_{Tc_1 \leq w \leq T c_2} \left( \frac{w+L}{T} \right)^{2L} \pr \big(\WKij = w\big) \nonumber \\
& \qquad {}+ \binom{K}{2} \pr\big( \WKij \notin [T c_1,T c_2] \big) \nonumber \\
& \leq \frac{K^2}{2(1-\epsilon_1)}\, \rev{\bigg(c_2 + \frac{\nu}{K}\bigg)^{2L}} \,\pr\big(  Tc_1 \leq \WKij \leq T c_2\big) \nonumber \\
&\quad + \binom{K}{2} \pr\big( \WKij < T c_1\big) + \binom{K}{2} \pr\big( \WKij > T c_2\big),
\end{align*} 
where we have used the fact that $L/T = \nu/K$.  This proves \eqref{eq:term2}.

\subsection{\rev{Proof of \eqref{eq:intersection}}} \label{sec:pf_intersection}

Recall that we condition on $\WKij = w$, and seek to bound the probability of $A_i \cap A_j$ for two defective items $i,j$. Here $A_i$ is the event that item $i$ is masked by the remaining defective items $\K \setminus \{i\}$ (one of which is $j$), and analogously for $A_j$.

In contrast to the rest of the paper, in this section, we represent the columns of the test matrix ${\mat X}$ corresponding to items $i$ and $j$ by lists $\mathcal{T}_{i}=(t_{i,1},\dotsc t_{i,L})$ and $\mathcal{T}_{j}=(t_{j,1},\dotsc t_{j,L})$. Each list entry is obtained by choosing $t\in\{1,\dotsc,T\}$ uniformly at random with replacement, so duplicates may occur.  Any given list occurs with probability $1/T^{L}$.

Without loss of generality, we assume that the $w$ tests containing items from $\K \setminus \{i,j\}$ are those indexed by $1,\dotsc,w$.  Letting $\mathcal{A}_i$ be the set of list pairs $(\mathcal{T}_i,\mathcal{T}_j)$ under which the event $A_i$ occurs, and similarly for $\mathcal{A}_j$, we have
\begin{equation}
    \Pr(A_{i} \cap A_{j} \,|\, \WKij = w) = \frac{N_{ij}}{T^{2L}}, \label{eq:pr_int0}
\end{equation}
where 
\begin{equation}
    N_{ij} = \sum_{\mathcal{T}_{i}}\sum_{\mathcal{T}_{j}}\II\big\{(\mathcal{T}_{i},\mathcal{T}_{j})\in \mathcal{A}_{i}\cap \mathcal{A}_{j}\big\} \label{eq:N_ij}
\end{equation}
is the number of pairs of lists in $\mathcal{A}_i \cap \mathcal{A}_j$.  Here the sets $\mathcal{A}_i$ and $\mathcal{A}_j$ implicitly depend on $w$.

To bound $N_{ij}$, we separately consider the number of `new positive tests' caused by items $i$ and $j$; that is, those not among the first $w$.  Specifically, letting $N_{ij}(\ell)$ be defined as in \eqref{eq:N_ij} with the summation limited to the case that there are $\ell$ such new positive tests, we have
\[
    N_{ij}=\sum_{\ell=0}^{L}N_{ij}(\ell).
\]
The summation only goes up to $L$ due to the fact that any new positive test containing $i$ must also contain $j$ and vice versa, otherwise the masking under consideration would not occur.

To bound $N_{ij}(\ell)$, we consider the following procedure for choosing the lists:
\begin{itemize}
    \item From $T-w$ tests, choose $\ell$ of them to be the new defective tests. This is one of $\binom{T-w}{\ell}$ options.
    \item For both $i$ and $j$, assign one list index from $\{1,\dotsc,L\}$
    to each of the $\ell$ new defective tests. This is at most $L^{\ell}$ options
    each, for $L^{2\ell}$ in total.
    \item For both $i$ and $j$, the remaining $L-\ell$ list entries are chosen
    arbitrarily from the $w+\ell$ positive tests. This is $(w+\ell)^{L-\ell}$
    options each, for $(w+\ell)^{2(L-\ell)}$ in total.
\end{itemize}
Combining these terms gives
\begin{align*}
    N_{ij}(\ell) & \le\binom{T-w}{\ell}\cdot L^{2\ell}\cdot(w+\ell)^{2(L-\ell)}\\
    & \le(T-w)^{\ell}\cdot L^{2\ell}\cdot(w+L)^{2(L-\ell)}\\
    & =(w+L)^{2L}\cdot\bigg(\frac{L^{2}(T-w)}{(w+L)^{2}}\bigg)^{\ell}.
\end{align*}
Now, since $w \ge c_1 T$ by assumption, and recalling that $T=\Theta(K\log N)$, that $K = \Theta(N^{\theta})$ with $\theta \in (0,1)$, and that $L = \nu T/K = \Theta(\log \rev{N})$, we find that the bracketed term is less than any fixed $\epsilon_1>0$ for sufficiently large \rev{$N$}. Hence, summing over $\ell$ gives
\begin{align*}
N_{ij} & \le\sum_{\ell=0}^{L}(w+L)^{2L}\cdot\bigg(\frac{L^{2}(T-w)}{(w+L)^{2}}\bigg)^{\ell}\\
& \le(w+L)^{2L}\sum_{\ell=0}^{\infty}\epsilon_1^{\ell}\\
& =(w+L)^{2L}\frac{1}{1-\epsilon_1}.
\end{align*}
Substituting into \eqref{eq:pr_int0}, we conclude that 
\[\Pr(A_{i}\cap A_{j} \,|\, \WKij = w) \le \bigg(\frac{w+L}{T}\bigg)^{2L}\cdot\frac{1}{1-\epsilon_1} , \]
as desired.

\subsection{\rev{Proof of Lemma \ref{lem:sss_set_vars}}}

\rev{We consider the procedure of selecting $T$ such that the fraction in the bracketed term in \eqref{eq:SSSfailure2a} equates to $1-\frac{1}{2(1-\epsilon_1)}$:
\begin{equation}
    1 - \frac{K (c_2 + \nu/K)^{2{\nu T/K}}}{ \vphantom{\hat1} 2c_1^{\nu T/K} (1-\epsilon_1) } = 1-\frac{1}{2(1-\epsilon_1)}, \label{eq:equate_half}
\end{equation}
or equivalently
\begin{equation}
    K \cdot \bigg(\frac{(c_2 + \nu/K)^2}{c_1}\bigg)^{\nu T/K} = 1. \label{eq:equate_half_a}
\end{equation}
Substituting $c_1 = 1 - \ee^{-\nu} - \sqrt{\frac{c_0 \ln T}{T}}$ and $c_2 = 1 - \ee^{-\nu} + \sqrt{\frac{c_0 \ln T}{T}}$} and performing some simple rearrangements, we obtain
\begin{align*}
\frac{\nu T}{K} &= \frac{ \ln K }{ \ln\frac{1 - \ee^{-\nu} - \sqrt{(c_0 \ln T)/T} }{ \big(1 - \ee^{-\nu} + \nu/K + \sqrt{(c_0 \ln T)/T}\big)^2 } } \\
&= \frac{ \ln K  }{ \ln\left(1 - \ee^{-\nu} - \sqrt{\frac{c_0 \ln T}{T}}\right) - 2 \ln\left(1 - \ee^{-\nu} + \frac{\nu}{K} + \sqrt{\frac{c_0 \ln T}{T}}\right) }.
\end{align*}
Applying Taylor expansions and noting that \rev{the terms $\sqrt{\frac{c_0 \ln T}{T}}$ and $\nu/K$} both decay as $O\big( \sqrt{\frac{\ln T}{T}} \big) = O\big( 1/\sqrt{K} \big)$, we obtain
\begin{equation}
\frac{\nu T}{K} = \frac{\ln K}{-\ln(1-\ee^{-\nu})} \bigg(1 + \rev{O\Big( \frac{1}{\sqrt{K}}\Big)} \bigg). \label{eq:L_choice}
\end{equation}
This choice is consistent with the assumed condition on $T$ in \eqref{eq:T_choice}.%, noting that 
%\[ K \ln K = \left(\frac{\theta}{1-\theta} K \ln\frac{N}{K}\right) \big(1+o(1)\big) \]
%when $K = \Theta(N^{\theta})$.

Having established \eqref{eq:L_choice}, we characterise the term $c_1^{\nu T/K}$ appearing in \eqref{eq:SSSfailure2}, with $c_1 = 1 - \ee^{-\nu} - \sqrt{\frac{c_0 \ln T}{T}}$ as above:
\begin{align*}
\ln( c_1^{\nu T/K} \big) 
&= \frac{\ln K \cdot \ln\bigg( 1 - \ee^{-\nu} - \sqrt{\frac{c_0 \ln T}{T}} \bigg)}{-\ln(1-\ee^{-\nu})} \bigg(1 + \rev{O\Big( \frac{1}{\sqrt{K}}\Big)} \bigg) \\
&= \big(-\ln K \big) \bigg(1 + \rev{O\Big( \frac{1}{\sqrt{K}}\Big)} \bigg),
\end{align*}
\rev{where we have again applied standard Taylor expansions.
Taking the exponential of  both sides gives $c_1^{\nu T/K} = 1/K^{1+O( 1/\sqrt{K})}$, which yields $K c_1^{\nu T/K} = 1 - o(1)$ since $K^{1/K^c} \to 1$ as $K \to \infty$ for any $c > 0$.  Substituting into \eqref{eq:SSSfailure2a} and recalling \eqref{eq:equate_half}, we obtain the desired bound \eqref{eq:SSSfailure3a}.}

%\begin{equation*}
%    \err{SSS} \geq K^{-\epsilon_6} (1 - \epsilon_4)\bigg( 1 - \frac{1}{2(1-\epsilon_1)} \bigg) - \epsilon_4. \label{eq:SSSfailure3}
%\end{equation*}

%\section*{Acknowledgments}

%M.~Aldridge was supported by the Heilbronn Institute for Mathematical Research.  % J.~Scarlett was supported by the `EPFL fellows' programme (Horizon2020 grant 665667).

\bibliographystyle{IEEEtran} 
\bibliography{bibliography}

%\newpage
%\input{response} %%% Alternatively, can put at the end. Or deleted entirely

\end{document}